# Review of Physics-based and Data-driven Multiscale Simulation Methods for Computational Fluid Dynamics and Nuclear Thermal Hydraulics

A Preprint

**Arsen S. Iskhakov***
Department of Nuclear Engineering
North Carolina State University
Campus Box 7909, Raleigh, NC 27695, USA
aiskhak@ncsu.edu

**Nam T. Dinh**
Department of Nuclear Engineering
North Carolina State University
Campus Box 7909, Raleigh, NC 27695, USA
ntdinh@ncsu.edu

## Abstract

Modeling of fluid flows requires corresponding adequate and effective approaches that would account for multiscale nature of the considered physics. Despite the tremendous growth of computational power in the past decades, modeling of fluid flows at engineering and system scales with a direct resolution of all scales is still infeasibly computationally expensive. As a result, several different physics-based methodologies were historically suggested in an attempt to "bridge" the existing scaling gaps.

In this paper, the origin of the scaling gaps in computational fluid dynamics and thermal hydraulics (with an emphasis on engineering scale applications) is discussed. The discussion is supplemented by a review, classification, and discussion of the physics-based multiscale modeling approaches. The classification is based on the existing in literature ones and distinguishes serial and concurrent approaches (based on the information flow between micro- and macro-models) with their variations.

Possible applications of the data-driven (machine learning and statistics-based) methods for enabling and / or improving multiscale modeling (bridging the scaling gaps) are reviewed. The introduced classification distinguishes error correction (hybrid) models; closure modeling (turbulence models, two-phase and thermal fluid closures); and approaches aimed at the facilitation of the concurrent physics-based multiscale modeling approaches. A comprehensive review of data-driven approaches for turbulence modeling is performed.

**Keywords** Review, Multiscale modeling, Data-driven closure model, Machine learning, Fluid dynamics, Thermal hydraulics

*Corresponding author

## 1. INTRODUCTION

Being a modeler, it is important to recognize that nature involves continuous interaction of multiple spatial and temporal scales – from the molecular level to small and large turbulent eddies and their interactions, and further to the influence of boundary conditions (BCs) of a considered domain. In fact, many practical applications in nuclear engineering are inherently multiscale and require corresponding adequate and effective modeling approaches that would accurately account for multiple spatial and temporal scales involved (Nourgaliev et al., 2012).

Despite the tremendous growth of computational power in the past decades, modeling of fluid flows for engineering scale problems with direct resolution of all scales is still infeasible due to prohibitively high computational cost (*e.g.*, it is impossible to use a molecular description to model coolant flow in a reactor core). As a result, a modeler must employ the divide and conquer approach. Instead of directly resolving of all scales, several successful simplifications[1] (scale separations, homogenization, filters, and / or averaging) should be introduced. These simplifications enable numerical modeling for engineering and scientific purposes: a computational analyst can focus on a "scale of interest" and consider the impact of other scales by providing necessary (missing) information for the developed model. One of the most popular approaches is to provide appropriate closure laws for governing equations. Theoretically, "ideal" and generic closure laws will not cause large errors in simulations. However, unfortunately, all closures have a limited range of applicability. This incorrect or missing physics can be referred to as "scaling gaps", which is caused by the inability to directly resolve all scales during modeling activities[2] (Bestion et al., 2012; Niceno et al., 2010).

Fig. 1.1 shows a pyramid that demonstrates different modeling approaches in computational fluid dynamics (CFD) and nuclear thermal hydraulics (TH).

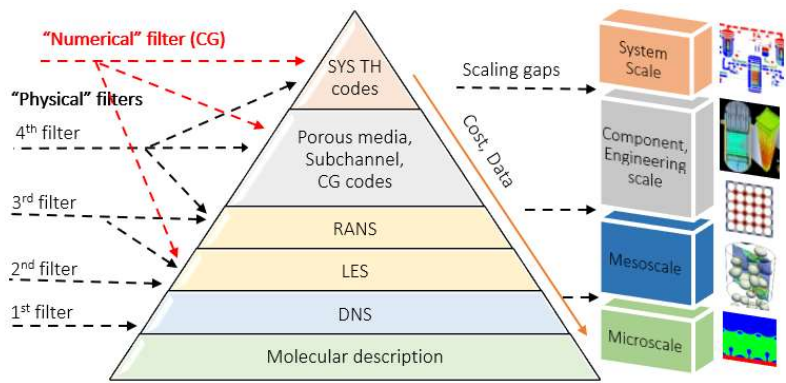

Fig. 1.1. Demonstration of scaling gaps in CFD and nuclear TH.

The bottom of the pyramid is a molecular description[3]. Again, due to the limited computational power, this approach can only be applied to a very limited number of problems at

---

[1] In some literature, these simplifications are referred to as "upscaling".
[2] This is a computational analysts' perspective (assumption: high-fidelity simulations, can provide ground truth results). From experimentalist's perspective scaling gaps are caused by the inability to perform full-scale experiments (assumption: full-scale experiments can provide ground truth values) (D'Auria & Galassi, 2010).
[3] Some of the methods for micro-scale modeling are reviewed in (Drikakis et al., 2019). Current work is focused on engineering scale simulations.

micro- and nano-scales. To model larger scales, it is necessary to apply 1st filter: average the behavior of molecules and introduce a concept of a continuous fluid with its macroscopic properties (*e.g.*, density ρ). Since here "a clear" scale separation is achievable (averaging of molecules is a good approximation), this simplification does not cause large errors in simulations. In this case the scaling gap is bridged by the equation of state (EOS). As a result, one can derive macroscopic conservation equations (the Navier-Stokes (NS) equations), which are considered as "exact" ones and written here for an incompressible fluid:

$$\frac{\partial u_i}{\partial x_i} = 0 \tag{1.1}$$

$$\frac{\partial u_i}{\partial t} + u_j \frac{\partial u_i}{\partial x_j} = -\frac{1}{\rho}\frac{\partial p}{\partial x_i} + \frac{1}{\rho}\frac{\partial \tau_{ij}}{\partial x_j} + g_i \tag{1.2}$$

where $u_i$ is velocity vector, $p$ is pressure, $x_i$ is coordinate vector, $t$ is time, $g_i$ is acceleration of a body force, $\tau_{ij}$ is stress tensor, which is assumed to be linearly correlated to the rate of strain $s_{ij}$ (for Newtonian fluids):

$$\tau_{ij} = \mu \cdot s_{ij} = \mu \cdot \left(\frac{\partial u_i}{\partial x_j} + \frac{\partial u_j}{\partial x_i}\right) \tag{1.3}$$

where μ is dynamic viscosity (an additional closure for μ should be provided to bridge the scaling gap). Eqs. (1.1) – (1.2) can be solved directly via numerical methods, which is referred to as direct numerical simulations (DNS). However, high requirements for mesh resolution for turbulent flows (Kolmogorov scales) result in high computational cost. As a result, DNS is only applicable to micro- and meso-scale problems (small scale phenomena such as energy cascade between turbulent eddies, bubble coalescence and breakup, *etc.*). One has to recognize that even DNS may require special multiscale / multiphysics treatment (*e.g.*, to take into account the roughness of a surface for BCs).

To reduce the computational cost, 2nd filter can be applied – field variables (velocity and pressure) are filtered for large eddy simulations (LES) or averaged for the Reynolds-averaged NS (RANS) simulations. This creates a turbulence closure problem: a model is required for subgrid scale (SGS) or Reynolds stress $R_{ij} = \langle u'_i u'_j \rangle$ in the governing equations:

$$\frac{\partial U_i}{\partial x_i} = 0 \tag{1.4}$$

$$\frac{\partial U_i}{\partial t} + U_j \frac{\partial U_i}{\partial x_j} = -\frac{1}{\rho}\frac{\partial P}{\partial x_i} + \frac{1}{\rho}\frac{\partial \tau_{ij}}{\partial x_j} + \frac{\partial R_{ij}}{\partial x_j} + \langle g_i \rangle \tag{1.5}$$

where $U_i$ is filtered / averaged velocity, $P$ is filtered / averaged pressure, symbol $\langle \ \rangle$ denotes here filtering / averaging operation. Among a dozen of developed turbulence models and their variations, none of them is applicable for all types of flows (Pope, 2000). Therefore, these models for LES and RANS simulations exhibit substantial model uncertainties, which should be appropriately addressed during the modeling efforts.

To decrease the complexity of the turbulence closure problem, often 3rd filter is applied – eddy viscosity hypothesis is employed for $R_{ij}$. The linear eddy viscosity hypothesis (the Boussinesq hypothesis) is conceptually similar to Eq. (1.3):

$$R_{ij} = -\langle u'_i u'_j \rangle = \nu_t \cdot \left( \frac{\partial U_i}{\partial x_j} + \frac{\partial U_j}{\partial x_i} \right) - \frac{2}{3} k \delta_{ij} \quad (1.6)$$

where $\nu_t$ is eddy (turbulent) viscosity, $u'_i$ is fluctuating component of velocity, $k = \frac{1}{2}\langle u'_i u'_i \rangle$ is turbulence kinetic energy (TKE), $\delta_{ij}$ is Kronecker delta. Such simplification introduces additional uncertainties in the modeling results. It has been shown that the linear eddy viscosity models are only applicable for "simple" flows (*e.g.*, without separations and sharp pressure gradients) (Xiao et al., 2016; Geneva & Zabaras, 2019). Nevertheless, LES and RANS simulations have noticeably lower computational cost than DNS, and they can be used for modeling of higher scales (up to the component scale, *e.g.*, 3D mixing effects in a reactor core, the effect of mixing vanes in fuel bundles, *etc.*) (Hohne et al., 2010).

When adiabatic and non-adiabatic two-phase flows are being modeled, 4$^{th}$ filter is often applied. For example, in a well-known two-fluid model, the behavior of bubbles / droplets is averaged: both the liquid and the gas phases are considered to be continuous, fully interpenetrating continua, coupled by interaction terms. These terms require closure models for interfacial and other interactions (Dennis, 2013). Thus, besides a turbulent closure problem, one must consider two-phase flow regimes, corresponding maps, and constitutive laws, which make this closure problem an order of magnitude more difficult than single-phase turbulence modeling (Nourgaliev et al., 2012).

Four discussed simplifications bring us to engineering and component scale codes (at best), which are dedicated to the design, safety, and operation studies for reactor components such as cores and heat exchangers (Liu et al., 2019; Wang et al., 2021). However, even with all the introduced "physical" filters, system scale simulations are infeasibly expensive. It is necessary to use an additional "numerical" filter – relatively coarse grids (CG), which places additional numerical and physical requirements for closure models. Mesh coarsening allows one to perform simulations at higher scales, *e.g.*, for system scale, which is dedicated to the overall description of a nuclear power plant or an integral test facility. The main applications of such system TH (SYS TH) codes are accidental transient simulations for safety analysis, operation studies, and real-time simulators (Bestion et al., 2012).

There are continuous efforts to improve the existing approaches for multiscale modeling. Some of them are physics-based (reviewed in Section 2), while others are data-driven (DD[4]) (reviewed in Section 3). Physics-based approaches rely on equation-based analysis of fluids. However, high dimensionality, nonlinearity, and multiscale nature of fluid flows defy closed-form solutions of the equations, which greatly undermines their usefulness. As will be discussed, DD approaches provide powerful information processing tools for better assimilation of existing computational and experimental data. They are flexible, not restricted by a specific form or equations, computationally efficient, and, therefore, have a promising potential to serve as another paradigm for multiscale bridging. In other words, DD methods can promise preservation of the "natural" continuous multiscale physics by a more efficient usage of big data by omitting activities related to compact (equation- and physics-based) representation of data (and unavoidable, data filtering and "wasting").

---

[4] In this work, the term "DD" is not only referred to machine learning (ML) techniques, but also includes all other tools and methods, which can effectively assimilate and analyze large amounts of data (*e.g.*, statistics-based methods, Bayesian inference, *etc.*). A DD approaches are "opposed" here to physics- (or equation-) based ones.

## 2. PHYSICS-BASED APPROACHES FOR MULTISCALE MODELING

### 2.1. Classification

Recent comprehensive review (Tong et al., 2019) introduces a classification of existing physics-based multiscale modeling approaches for fluid flow and heat transfer problems. The classification is performed based on the available knowledge about macro-models[5]. In this work, this classification is adopted in a slightly modified version and presented in Fig. 2.1. Other classifications are available in literature, *e.g.*, (Nourgaliev et al., 2012; Ingram et al., 2004; Scheibe, et al., 2015), which are not contradictory, but rather complementary ones.

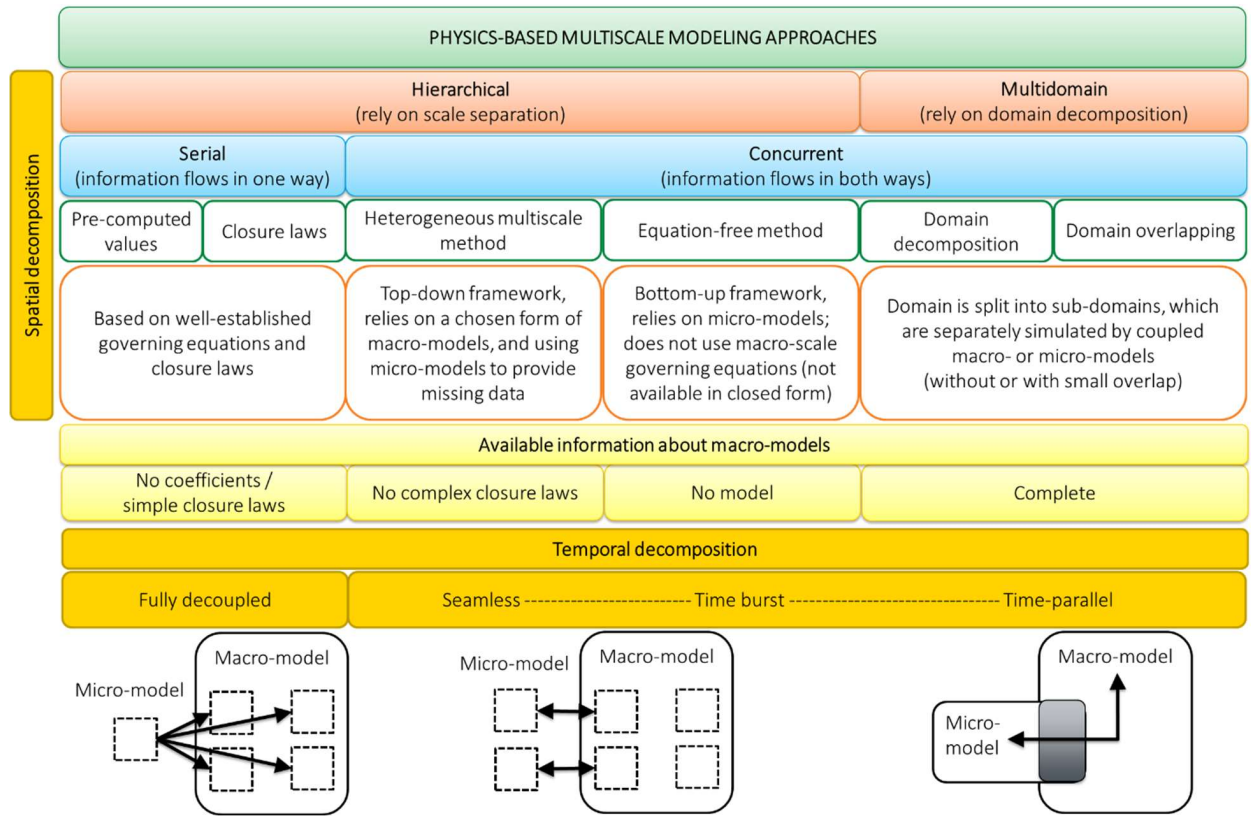

Fig. 2.1. Classification of physics-based multiscale modeling approaches.

First, from the spatial decomposition perspective, hierarchical and multidomain approaches are distinguished:

(i) **Hierarchical approaches** rely on separation of scales in the computational domain. An illustrative example can be a framework, which employs micro- and macro-models. Each model is being used for simulation of the whole domain and resolves its own scale. The information exchange between these scales occurs at some time moments depending on the adopted coupling scheme.

(ii) **Multidomain approaches** rely on a decomposition of the computational domain into sub-domains. Each sub-domain is assigned to either a micro- or a macro-model without (**domain**

---
[5] Further "micro-" and "macro-models" are simply referred as synonyms for smaller and larger scale models.

**decomposition**) or with small overlapping (**domain overlapping**) (Zhang, 2020). The models are matched either over a handshake region or across an interface, by properly filtering solution data. A framework based on two coupled codes can serve as an example for such approaches (*e.g.*, a SYS TH code for a whole plant and a RANS solver for a reactor core). The multidomain approaches can be used when complete micro- and macro-models are available. No time decomposition is employed, and (most often) computation goes on in parallel with "continuous" data exchange between the sub-domains.

Next, it is recognized that in the hierarchical approaches, information can flow either in one direction (*e.g.*, from a micro- to a macro-model), or it can flow in both ways. Thus, **serial** and **concurrent** families are distinguished here. Serial approaches can use either **pre-computed** values (*e.g.*, tables, constants, *etc.*) or **closure laws**. The missing information about the macro-model (well-established governing equations) can be either coefficients / constants or relatively simple[6] closure laws. From the temporal decomposition perspective, serial approaches are fully decoupled – pre-computed values and closure laws are developed separately from the governing equations. The hierarchical concurrent approaches can be largely subdivided into **heterogeneous multiscale method** (HMM) and **equation-free method** (EFM). The HMM (E et al., 2009) can be characterized as a top-down framework, relying on a chosen form of a macro-model, and using a micro-model for providing missing data to advance the macro-model in time. It is applicable when there are governing equations available for the macro-model, but they lack "complex" closure laws (Nourgaliev et al., 2012). The EFM (Gear et al., 2003; Li et al., 2007; Kevrekidis, et al., 2003) enables modeling of extended spatiotemporal scales using appropriately initialized micro-model (bottom-up framework). It does not use explicit macro-scale governing equations since they are not available in closed form.

For the temporal coupling, besides the parallel computation, the **seamless** (mostly used for the HMM) or **time burst** (mostly used for the EFM) schemes can be adopted. In the time burst scheme, the macro-scale information is extracted from short bursts of the micro-scale simulations. In contrast, the seamless method does not require reinitializing the micro-model at each macro-timestep. Instead, the macro- and micro-models evolve simultaneously using different time steps (and therefore different clocks), and exchange data at every step. The micro-model uses its own appropriate time step, while the macro-model runs at a slower pace than required by accuracy and stability considerations, to allow the micro-model to relax (E et al., 2009).

Such important multiscale methodologies as **multigrid methods** and **mesh and algorithm refinement approaches** are not included in the classification, but they will be discussed later. They can be categorized as serial approaches with "algorithmically" reduced computational cost.

Further improvements of physics-based multiscale modeling can be achieved by improving the understanding of physics and, subsequently, mathematical models. This is enabled by the **multiscale analysis**, when micro- and macro-scale simulations or experiments are separately used to understand the physics, test hypotheses, and / or develop better closure models. Additionally, further advances of the coupling methodologies and algorithms can be beneficial for the concurrent modeling approaches (Bestion et al., 2012; Niceno et al., 2010).

## 2.2. Serial Modeling

Even though, there are several examples of using pre-computed values (*e.g.*, form loss coefficients, fluid properties, *etc.*), there is little hope that pre-computation can become practical for many nuclear reactor TH problems since the parameter space is impractically large. Therefore, multiscale modeling should be based whether on closure modeling, or on concurrent coupling (Karolius & Preisig, 2018) linking together micro- and macro-models "on-the-fly" as the computation goes on (Nourgaliev et al., 2012).

---

[6] "Simple" here denotes availability of a closure model and its validity for the intended use.

Historically, closure modeling is the most popular approach to consider different scales simultaneously. Unfortunately, it has limited capabilities since it is nearly impossible to develop generic closures that would work in all circumstances. In relation to nuclear TH, closure modeling can be subdivided into:
- closures for macroscopic properties, such as EOS. As discussed in the introduction, since $1^{st}$ filter allows to achieve a "clear" scale separation, equation-based and tabulated macroscopic properties lead to accurate solutions. Even though it is not easy to obtain an EOS, this problem will not be discussed in this work.
- turbulence models.
- two-phase and thermal fluid closures.

### 2.2.1. Turbulence models

Among the most popular approaches for turbulent flows modeling DNS, LES, and RANS simulations can be distinguished (Pope, 2000). DNS does not require turbulence closures since it resolves all necessary scales directly. As a result, this approach is computationally expensive. A comprehensive review of traditional turbulence models for LES and RANS goes far beyond the scope of this work; here only high-level classification is provided, see Fig. 2.2.

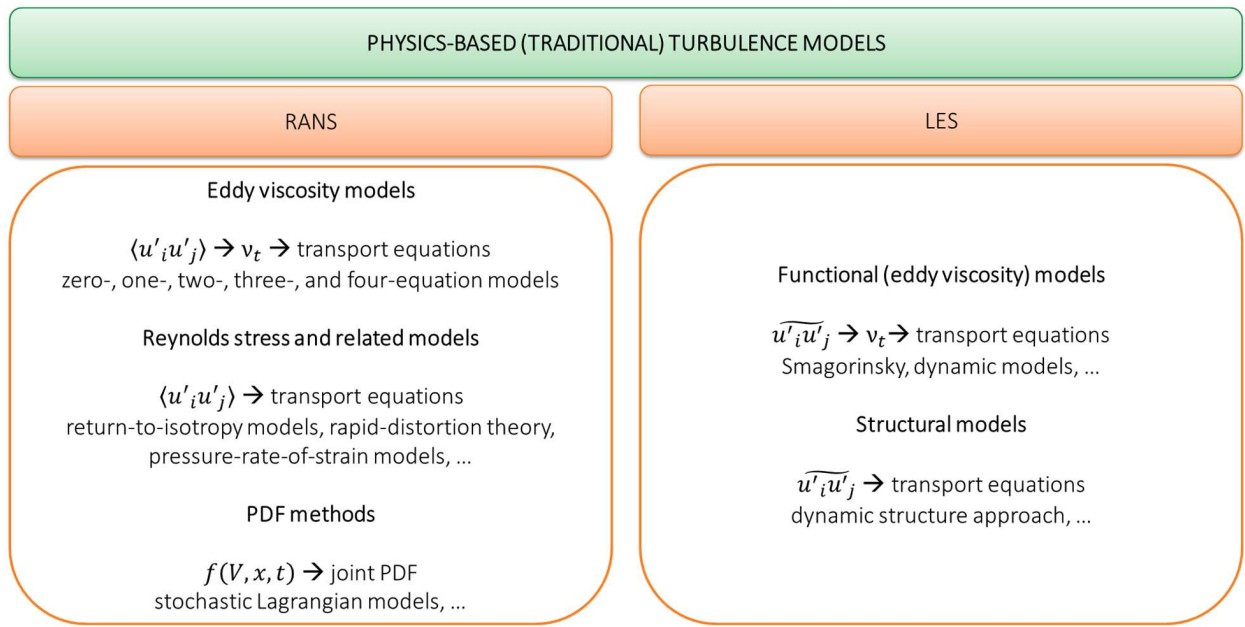

Fig. 2.2. Classification of physics-based (traditional) turbulence models for RANS and LES.

RANS turbulence models can be classified into three groups (Pope, 2000; Iaccarino, 2020):
- models that employ the eddy viscosity hypothesis Eq. (1.6) (**turbulent viscosity models**). By turn, such models can be classified according to the number of equations employed for such quantities as turbulent viscosity, TKE, *etc.*: zero-equation (*e.g.*, mixing-length model), one-equation (*e.g.*, Spalart-Allmaras model), two-equation (*e.g.*, $k-\varepsilon$ model) three-equation (*e.g.*, $k-\varepsilon-A$ model), and four-equation models (*e.g.*, $v^2-f$ model).
- models that are based on the transport equations for the Reynolds stress (less often algebraic (Drikakis et al., 2019)) (**Reynolds stress and related models**). Examples include return-to-isotropy models, rapid-distortion theory, pressure-rate-of-strain models, *etc.*
- **probability density function (PDF) methods** (Pope, 1985). The central idea of these methods is solution of stochastic transport equations for joint PDF $f(V, x, t)$, where $V$ is sample space variable to velocity.

Drawbacks and ranges of applicability of RANS turbulence models are discussed in detail in (Pope, 2000; Xiao & Cinnella, 2019) and other literature. Here we just note that the models perform best in certain flow conditions / geometry / regimes at quasi-steady state (*i.e.*, a clear separation between the mean flow and stochastic components is assumed); see *e.g.*, (Launder & Spalding, 1974; Wilcox, 1998). However, uncertainty significantly increases when they are applied to describe long transient TH scenarios, during which the flow regime over the computational domain varies greatly (especially when a flow exhibits transitional from laminar to turbulent behavior). In other words, there is no a universal turbulence model that would work for all cases, but only recommendations collected from their applications and assumptions used for models' derivation.

LES for turbulent flows is a less popular (but more promising) approach than RANS modeling. Due to this, there are less turbulence models developed, and less knowledge is accumulated. Similarly, these models can be categorized into **eddy viscosity models** (*e.g.*, Smagorinsky, dynamic models, *etc.*) and **structural models** (*e.g.*, dynamic structure approach (Rutland, 2004)). Eddy viscosity approaches are more popular and methodologically similar to the turbulent viscosity models for RANS, while structural models are analogous to Reynolds stress models for RANS since they employ transport equations for the SGS stress.

Among the challenges and issues of LES, following can be emphasized. As RANS, LES require further developments of SGS models (current ones are proven to be accurate only for specific conditions). Specification of inflow BCs is a challenging problem for LES since it should include information on the turbulence properties. Accurate wall layer modeling makes LES very computationally expensive approaching cost of DNS (Zhiyin, 2014). Thus, there are also hybrid RANS-LES models available to avoid this problem (*e.g.*, detached eddy simulations, when RANS is used in the near wall regions, while LES is used for the bulk); additionally, (physically) constrained LES approaches are under investigation for SGS stress modeling (Chen et al., 2014).

*2.2.2. Two-phase flow and thermal fluid closures*

Without loss of generality, let us consider the two-fluid model (Two-fluid model, 2020) which is based on the solution of two sets (for each phase) of governing equations for conservation of mass, momentum, and energy. Fig. 2.3 shows a typical structure and interaction of closure models required for a two-fluid model solver (Liu et al., 2019).

Additionally to a turbulence model, it is also necessary to provide closures for boiling / convective heat transfer, nucleation, interfacial mass and momentum exchange, flow regime maps, *etc.* Thus, two-phase flow and thermal fluid closure problem is several orders of magnitude a more difficult problem than modeling of turbulent single-phase flows. Hence, a comprehensive review goes beyond the scope of this work. Perhaps, each of the closure models requires a separate review to get a good idea of their ranges of applicability and deficiencies. We note that similarly to the RANS turbulence models, two-phase and thermal fluid closures are designed to work in certain flow conditions / geometry / regimes at quasi-steady state. This is caused by the development framework of such closures – as a rule they are obtained in steady-state separate effect test facilities at certain conditions (often with a reduced scale). The paradox is that they are mostly used to model transient processes. As a result, uncertainties of such modeling are relatively large.

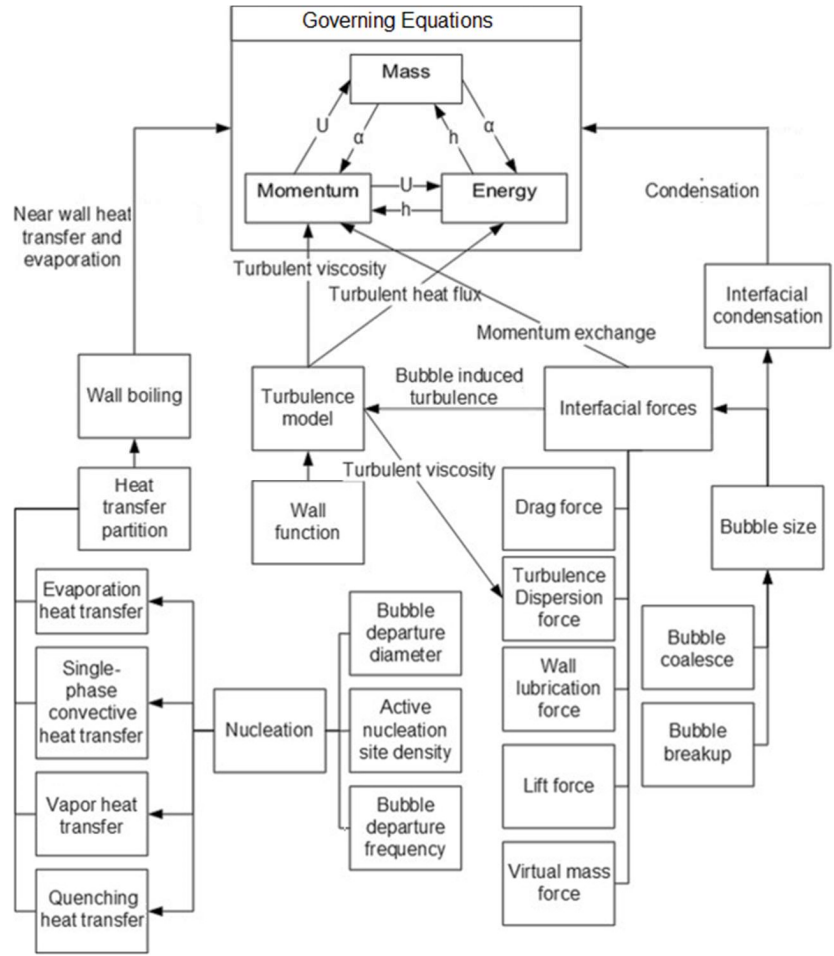

Fig. 2.3. Closure models in a two-fluid model solver (reproduced from (Liu et al., 2019)).

Historically, many semi-empirical correlations for thermal fluid modeling were derived with help of the scaling analysis (dimensionless analysis and dimensionless governing equations) (D'Auria & Galassi, 2010; Bestion, 2017; Dzodzo, 2019). The scaling analysis is historically being used for addressing "experimental" scaling gaps, when it is necessary to extrapolate results obtained on reduced-scale integral and separate effect test facilities (*e.g.*, volume scaling, three-level scaling, power-to-mass scaling, *etc.*). Despite a notable success in SYS TH analysis, scaling analysis has a limited potential for bridging "numerical" scaling gaps due to inevitable distortions, which lead to large uncertainties (acceptable for SYS TH scale, but unacceptable for fine-scale physics). Though, some examples of successful using of the scaling analysis for extrapolation of results of numerical simulations can be found in literature (for instance, based on a small-scale CFD simulations and dimensionless analysis, Ciuffini et al. (2016) predicted heat exchanger performance).

### 2.3. Concurrent Modeling

#### 2.3.1. Multidomain approaches

Review of coupled codes for nuclear TH applications is performed in (Zhang, 2020). The paper gives a good historical retrospective as well as provides specific examples of coupling efforts (*e.g.*, TRAC/COBRA-TF, RELAP5/COBRA-TF; RELAP/FLUENT, ATHLET/CFX, *etc.*). It is noted that the coupling of RANS and SYS TH codes is now the main trend in the nuclear community. Nevertheless, the coupling of subchannel and SYS TH codes is also under consistent investigation. As a supplementation, the coupling of subchannel codes and RANS codes is also drawing attention nowadays though there are limited papers published.

Zhang (2020) classified and assessed coupling approaches from the computer science (not provided) and mathematical perspective (see Fig. 2.4). The spatial coupling issue arises from the different nodalizations in coupled codes. It can be either domain coupling (includes domain decomposition and domain overlapping) or mesh / field coupling (translation of mesh from one to another via manually defined relations, user-developed subroutine, or 3$^{rd}$ party library). The difference between domain decomposition and domain overlapping (Grunloh, 2016) can be explained by a following example. For the domain overlapping, a SYS TH code models an entire plant, while a RANS code simulates only a small overlapping part. In the domain decomposition, both codes simulate only their parts contacting through an appropriate interface (BCs). Temporal coupling is subdivided into weak (explicit and semi-implicit) and strong (implicit) couplings.

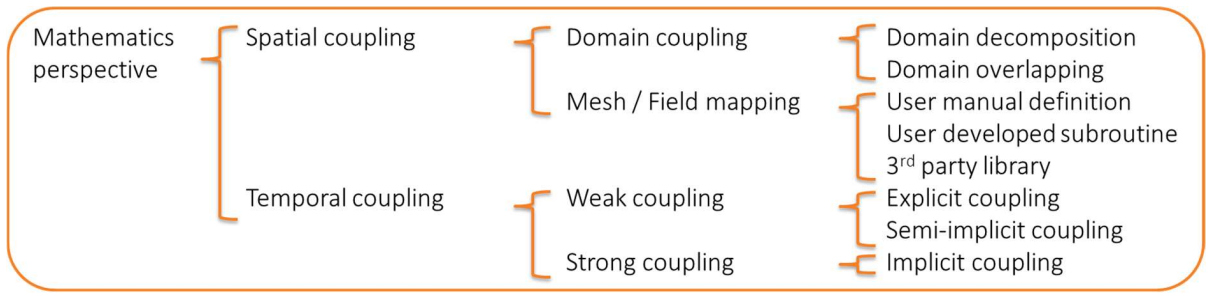

Fig. 2.4. Classification of the multidomain approaches from the mathematical perspective (reproduced from (Zhang, 2020)).

There are several projects aimed at improving multiscale simulations for nuclear TH. European projects NURESIM and NURISP (Bestion et al., 2012) were aimed at developing and validating a multiphysics and multiscale platform for nuclear reactor simulations. The platform includes SYS TH, subchannel, and CFD (DNS, LES, and RANS) simulation tools. Micro-scale simulations were used for a better understanding of physical phenomena, for the prediction of small-scale effects, and for simulation of problems that require a fine space and / or time resolution. Thus, pressurized thermal shock is investigated using DNS, LES, and RANS approaches, and then a coupling with a SYS TH is performed. Condensation-induced water-hammer is also investigated at both CFD and SYS TH levels. Boiling flow in a reactor core up to departure from nucleate boiling (DNB) or dry-out is investigated at scales much smaller than classical subchannel analysis codes. DNS is used to investigate very fine processes, while RANS and LES were employed to simulate bubbly and annular mist flow. Loss of coolant accidents (LOCA), which are usually treated by SYS TH codes were also revisited at the CFD scale.

MSMA project (Niceno et al., 2010) was aimed at formulation of improved closure laws for CFD prediction of convective boiling and critical heat flux (CHF). The authors distinguished four scales of interest: (1) macro-scale (a few centimeters) to model TH phenomena within one or more subchannels using RANS or LES; (2) meso-scale (millimeters) to model bubbles growing and detaching from the wall using DNS; (3) theoretical approach for micro-scale (micrometers) effects investigation (movement of the triple interline during the bubble life cycle and the evaporation of the thin liquid film); (4) nano-scale (nanometers; molecular scale) which is useful to analyze processes at the interface between vapor and liquid, or at the triple contact line between solid vapor and liquid (molecular dynamics modeling). To link all the scales, the authors relied on the multidomain approach and multiscale analysis (separate use of codes).

Similar activities are performed in CASL project (CASL, 2013). Challenges addressed include better prediction of CRUD phenomenon, pellet-cladding interaction, cladding integrity, DNB, flow regimes transition, and fretting. As can be seen, the focus was placed on multiphysics modeling (which is even more challenging than multiscale TH), however, some of the challenges required multiscale TH treatment and employment of codes with different scale resolution capabilities (DNS, RANS, *etc.*).

### 2.3.2. Heterogeneous multiscale method

The HMM is a general top-down framework that relies on a macro-model and uses micro-model to provide missing data. It is promising for nuclear TH applications, since usually decent macro-models are available (*i.e.*, conservation equations) and generic and reliable closures are missed (Nourgaliev et al., 2012). The name "heterogeneous" is used to emphasize that the models at different scales may be of very different nature, *e.g.*, molecular dynamics at the micro-scale and continuum mechanics at the macro-scale (E et al., 2007).

The HMM can be explained as follows. Consider a macro- and a micro-model:

$$\frac{\partial U(t)}{\partial t} = \mathcal{F}(U, \mathcal{D}) \qquad (2.3.1)$$

$$\frac{\partial u(t)}{\partial t} = f(u, d) \qquad (2.3.2)$$

where $U$ and $u$ are macro- and micro-scale state variables, $\mathcal{D}$ are data required to close the macro-model (*e.g.*, Reynolds stress, interfacial drag, *etc.*), $d$ are data needed to update the micro-model (*e.g.*, set of constraints, initial and BCs).

The goal of the HMM is to compute the evolution of $U$ using the assumed form of $\mathcal{F}$ coupled together with the micro-model. The primary purpose of the micro-model is to provide missing macro-scale data $\mathcal{D}$, using an appropriate data processing procedure (Nourgaliev et al., 2012). Macro- and micro-scale variables are connected by compression ($\mathcal{C}$) and reconstruction ($\mathcal{R}$) operators, which must be consistent with each other:

$$\mathcal{C}u = U \qquad \mathcal{R}U = u \qquad \mathcal{CR} = \mathcal{I} \qquad (2.3.3)$$

where $\mathcal{I}$ is the identity operator. A good $\mathcal{C}$ and $\mathcal{R}$ should be physically meaningful, mathematically stable, computationally efficient, and easy to be implemented (Tong et al., 2019). However, one cannot expect the two operators are commutative and usually $\mathcal{CR} \neq \mathcal{I}$ because the micro-scale information is filtered in $\mathcal{R}$. For a particular application, there is a significant amount of work to implement the framework, such as designing $\mathcal{C}$ and $\mathcal{R}$, constraints formulation, data estimation, initialization of micro-solvers, *etc.*

Tong et al. (2019) furtherly distinguish analytical (asymptotic analysis / homogenization) and numerical HMM. The basic idea of homogenization is to expand the considered variables such as temperature and velocity into the macroscopic averaged variables and microscopic fluctuation variables. By the multiscale analyses of the governing equation, a homogenized equation for the macroscopic averaged variable and a cell problem in the representative elementary volume for the microscopic fluctuation variables can be derived. In the numerical HMM the processes are usually described by the governing equations or numerical methods at different scales. As a result, they do not have analytical connection.

General review of the methodology and several examples of application HMM to different problems is given in (E et al., 2007). It is worth noting that most papers are not focused on engineering scale problems (*e.g.*, atomistic scale is considered by micro-models; geomechanical scale is considered by macro-models). Examples of applications for fluid flows modeling include the study of transient unsaturated (Chen & Ren, 2008) and saturated water flow (Chen & Ren, 2014), the study of single-phase flow in porous media (Chu et al., 2012; Alyaev et al., 2014).

*2.3.3. Equation-free method*

The main idea of the EFM is to extract macro-scale information from short bursts of micro-scale simulations in space and time (bottom-up approach). The EFM consists of following steps: coarse time stepper, coarse projective integration, gap-tooth scheme, and patch dynamics. In the EFM the compression and reconstruction operators are usually called restriction and lifting (Tong et al., 2019).

During the coarse time stepper, the macro-variables evolve for a micro-timestep $\delta t$. The procedure consists of lifting, microscale simulation, and restriction steps:

$$\begin{aligned}
\mathcal{R}U(X,t) &\to u(x,t), d(x,t) \\
u(x,t+\delta t) &= u(x,t) + \delta t \cdot f(u,d) \\
\mathcal{C}u(x,t+\delta t) &\to U(X,t+\delta t)
\end{aligned} \quad (2.3.4)$$

The coarse projective integration projects the macro-scale evolution to larger macro-timestep $\Delta t$. The micro-model evolves for $M \cdot \delta t$, which is larger than the relaxation time, so that the effects of the initial artificial higher-order moments can be eliminated. The changing rate of macroscopic variables is approximated by:

$$\mathcal{F}\big(U(X,t)\big) \approx \frac{U(X, t + M \cdot \delta t) - U(X,t)}{M \cdot \delta t} \quad (2.3.5)$$

therefore, using temporal extrapolation macroscopic evolution is

$$U(X, t + \Delta t) = U(X, t + M \cdot \delta t) + (\Delta t - M \cdot \delta t) \cdot \mathcal{F}\big(U(X,t)\big) \quad (2.3.6)$$

The gap-tooth scheme is proposed for the scale separation. The space is divided into control volumes located at $x_i$; size of these control volumes is $[x_i - h/2, x_i + h/2]$, where $h$ is much smaller than the size of a macro-control volume, but large enough for the micro-model to eliminate the artificial effects of BCs. Then the gap-tooth scheme consists of lifting, microscale simulation, and restriction steps:

$$\begin{aligned}
\mathcal{R}U(x_i \pm h/2, t) &\to u(\xi_i, t), d(x_i \pm h/2, t) \\
u(\xi_i, t + \delta t) &= u(\xi_i, t) + \delta t \cdot f\big(u(\xi_i, t), d(x_i \pm h/2, t)\big)
\end{aligned} \quad (2.3.7)$$

$$\mathcal{C}u(\xi_i, t + \delta t) \to U(x_i, t + \delta t)$$

where $\xi_i$ are spatial coordinates of the micro-model.

The patch dynamics is the combination of Eqs. (2.3.6) and (2.3.7). It is used to extend the microscopic simulations in small spatial and temporal steps into the macroscopic large spatial and temporal steps.

Bottom-top strategy seems to be very attractive when there is limited knowledge on macro-models and when scale separation is clearly defined. Therefore, they are useful for "upscaling" from quantum mechanical or molecular scales, but hardly practical for nuclear TH applications.

*2.3.4. Time coupling schemes*

Previous discussion of the multidomain approaches, HMM, and EFM was mostly focused on the spatial coupling, while there are several opportunities for temporal coupling. Following (Tong et al., 2019), time parallel scheme, time burst scheme, and seamless approach are distinguished.

Time parallel schemes are employed when there is no significant temporal scale separation. Both macro- and micro-models evolve simultaneously in time. The data exchange occurs at specific time steps (*e.g.*, each macro-timestep $\Delta t = N \cdot \delta t$). The computational cost of the micro-model is not reduced since it performs simulation during the whole transient.

Time burst scheme can be applied if there is a separation of time scales and the relaxation time is much smaller than $\Delta t$. The micro-model evolves for $M \cdot \delta t < \Delta t$ (should be large enough for the micro-model to reach a quasi-equilibrium state). In the HMM the micro-model evolves under the constraint of the macro-model and then provides the missing data to the macro-model. In the EFM the coarse projective integration extrapolates the short time step to longer time step. Potentially, the time burst scheme can be also adopted in the multidomain methods.

While in the time burst scheme the micro-model should be reinitialized before every burst (which is computationally expensive (E et al., 2009).), in the seamless method the micro-model evolves continuously without reinitialization. Like in the time burst scheme, the micro-model needs to evolve for $N \cdot \delta t$ to achieve a quasi-equilibrium state. The difference is that the macro-timestep $\Delta t$ may not to be consistent with $N \cdot \delta t$, and it can be a much larger $\Delta t = M \cdot N \cdot \delta t$. Therefore, the scheme can be regarded as that the macro-model drives the micro-model.

*2.3.5. Adaptive mesh & algorithm refinement and multigrid methods*

Other, often employed, but not discussed methods that enable multiscale modeling include:
- adaptive mesh and algorithm (model) refinement, in which mesh or / and model are adapting during the solution (mesh is becoming finer and / or a high-fidelity model is being used in regions where micro-scale effects are important, and *vice versa*) (Berger & Colella, 1989; Garcia et al., 1999).
- multigrid methods, which are based on classical relaxation schemes (that are generally slow to converge but fast to smooth the error function) and approximating the smooth error on a CG, by solving equations which are derived from partial differential equations (PDE) and from the fine-grid residuals. The solution of these CG equations is obtained by using recursively the same two processes (Ghia et al., 1982; Fulton et al., 1986).

These methods do not employ strong spatial scale separation comparing to the HMM and the EFM[7].

The idea of the adaptive mesh and algorithm (model) refinement is natural. Computational domain usually consists of regions with micro-scale effects that require accurate solution (such as boundary layers, shock front, near bubble regions, *etc.*) and regions where micro-scale effect are not as important (*e.g.*, large enclosures, tube outlets, *etc.*). To reduce the cost of simulations, first idea would be using a fine mesh for micro-scale phenomena regions, and a CG for regions with macro-scale phenomena. Similarly, if a computationally expensive closure model can be replaced by a simpler model for regions with macro-physics, this can significantly reduce computational cost and, therefore, allow multiscale treatment. Such adaptation can be either automatic (*e.g.*, mesh is becoming finer if discretization error increases), or manual (based on the knowledge of physics). There are several examples of using such approaches in CFD and TH, see *e.g.*, (Garcia et al., 1999; Berger & Colella, 1989; Cook, 1999). Perhaps, the disadvantage of this approach is in the lack of scalability: too CG and too "simple" models can lead to significant errors in macro-scale regions and "pollute" solution in micro-scale regions.

The main idea of the multigrid methods is to accelerate the convergence of iterative methods by a global correction of the fine grid solution approximation from time to time, accomplished by solving a coarse problem. This recursive process is repeated until a grid is reached on which the cost of direct solution is negligible compared to the cost of one relaxation sweep on the fine grid (Trottenberg et al., 2001). Despite a well-developed mathematical apparatus, multigrid methods are also hardly scalable for engineering and system scale problems since they work best for elliptic problems (*e.g.*, steady state flows), while the NS equations are hyperbolic in time. As a result, a long transient with high mesh resolution (a typical requirement) is a computationally expensive exercise for the multigrid methods. The HMM can be considered as an extension of the multigrid method (Brandt, 1977; 2003; 2009) since ideologically it relies on similar operations.

*2.3.6. Multiscale Analysis Platform*

Previous discussion of the multiscale modeling approaches is a high-level introduction. There are several problem specific variations of such methods, which are not covered by this review. A multiscale analysis platform introduced in (Scheibe, et al., 2015) provides a guidance for choosing a specific methodology for a problem of interest. The platform is presented in Fig. 2.5. By asking the questions about the degree of coupling, the spatial and temporal scale separation, the size of the micro-scale region and the knowledge about the macro-model, the platform leads to a specific motif.

Thus, motif A is mainly composed by the direct resolution of all scales (*e.g.*, DNS, multigrid and adaptive mesh refinement methods). Motif B leads to the closure modeling approaches (*e.g.*, RANS). If the answer for the question 2 is "micro- and macro-models are tightly coupled", then one arrives to the variations of HMM (motifs E1 and F) and the EFM (motifs E2 and G) depending on the temporal and spatial scale separations. If temporal scale separation is unachievable, then one arrives at motif H with the time-parallel multidomain approaches (coupling of codes). More detailed discussion with specific examples can be found in the original source (Scheibe, et al., 2015).

---

[7] Brandt (2009) argues otherwise: "the HMM does not employ substantial scale separation". Perhaps, the controversy arises due to the subjectiveness of such words as "strong", "substantial".

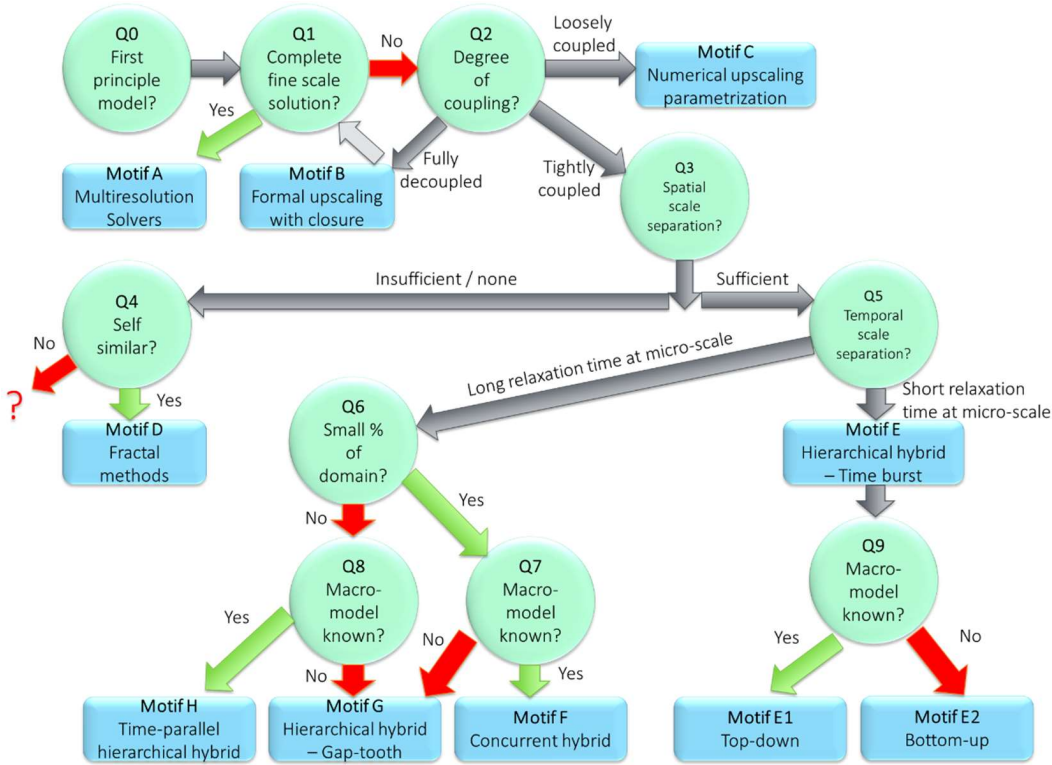

Fig. 2.5. Multiscale analysis platform (reproduced from (Scheibe, et al., 2015)).

### 3. DATA-DRIVEN APPROACHES FOR MULTISCALE MODELING

#### 3.1. Classification

Literature review has shown that DD approaches to enable and / or improve multiscale modeling in CFD and nuclear TH fields can be classified into four categories depending on a goal for which a DD application is employed (Fig. 3.1):

(i) **Error correction in physics-based models** (also known as hybrid methodology). In such approaches a DD application works together with a physics-based model. Therefore, it is not an alternative, but rather a complementary tool for a physics-based closure. The main goal is to correct errors in physics-based models before, during, or / and after simulations.
(ii) **Turbulence models**. Here a DD application comprises a "full" turbulence model to account for SGS turbulent effects in a flow. Therefore, such turbulent closures serve as an alternative approach for traditional turbulence models.
(iii) **Two-phase and thermal fluid closures**. Similarly to (ii), a DD application serves as an alternative approach to a physics-based closure for modeling of interfacial forces, heat transfer and other effects.
(iv) **Facilitation of concurrent multiscale modeling**. As discussed in Section 2, concurrent multiscale modeling consists of the HMM, the EFM, and the multidomain approaches. The last one heavily relies on the algorithmic coupling of existing codes, which is the most problematic issue. The HMM and the EFM are based on the mapping (going back and forth between the scales) of micro- and macro-models. There is a high potential for DD approaches to reduce the complexity of such modeling, which will be discussed later.

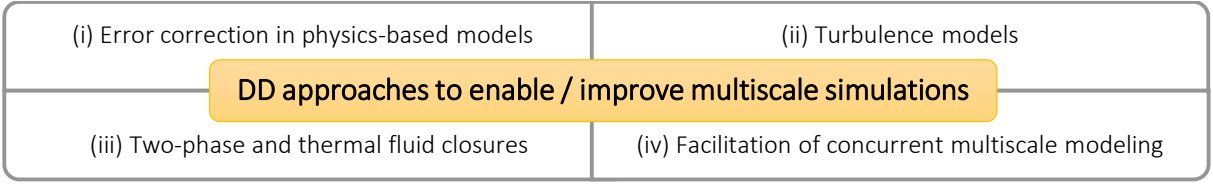

Fig. 3.1. DD approaches to enable / improve multiscale simulations in nuclear TH and CFD.

Approaches (i), (ii), and (iii) can be furtherly subdivided into: (a) **high-to-low** methodology (Lewis et al., 2016) (informing low-fidelity simulations by high-fidelity data, *e.g.*, use DNS data to improve a turbulence model for RANS), and (b) **computational cost reduction** of high-fidelity simulations to make them applicable for modeling of higher scales (*e.g.*, decrease the computational cost of DNS to model an engineering scale problem). This classification is illustrated by Fig. 3.2.

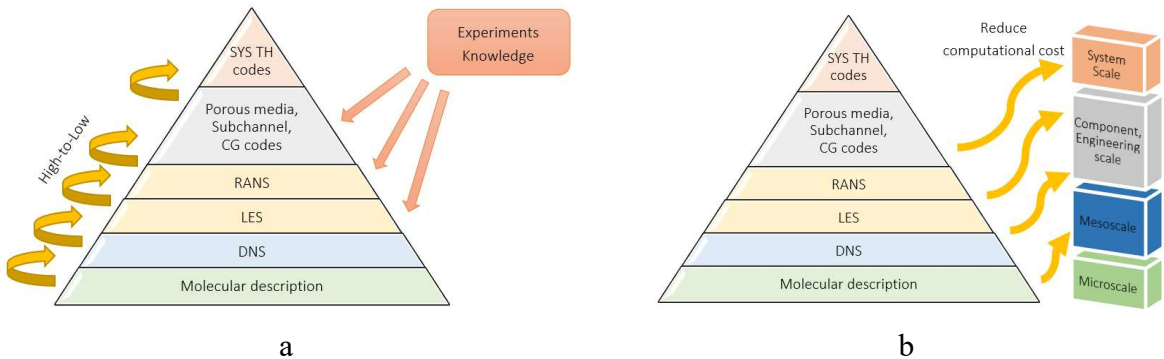

Fig. 3.2. (a) High-to-low and (b) computational cost reduction approaches.

### 3.2. Error Correction in Physics-based Models

Classification of DD error correction methods is illustrated in Fig. 3.3. Among high-to-low approaches, following can be distinguished from the "scale bridging" point of view: (1) informing LES by DNS; (2) informing RANS simulations by DNS / LES / knowledgebase, and / or experimental data; (3) informing CG or SYS TH simulations by RANS simulation or experimental data. Computational cost reduction approaches can be subdivided into (1) grid-coarsening methods; (2) grid-optimization methods; (3) initial approximation and co-simulation methods.

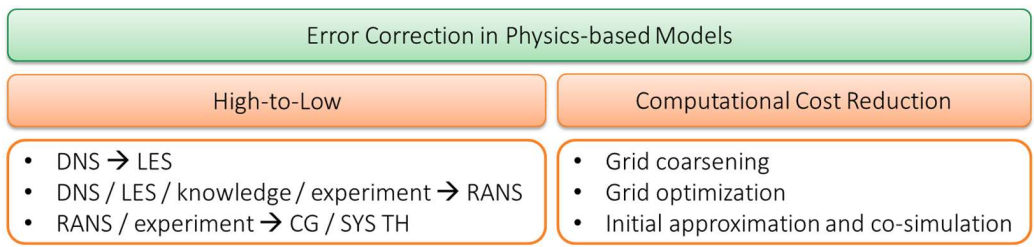

Fig. 3.3. Classification of DD approaches for error correction in physics-based models.

*3.2.1. High-to-low approaches*

**DNS to LES**. With the availability of high-fidelity data from DNS it seems natural to use them to inform codes with lower fidelity such as LES. Such approaches can be considered as attempts to bridge a gap caused by filtering of the NS equations to obtain governing equations for LES. As a result, LES require a SGS stress model.

Beck et al. (2019) used CG (filtered) DNS data to improve LES for a decaying homogeneous isotropic turbulence. Spatial terms in the filtered NS equations are predicted by a residual NN (special architecture of a convolutional NN (CNN) with linear skip connections, which improve prediction accuracy without overfitting issues (He et al., 2016)). The framework allowed to obtain "perfect" filtered momentum equation:

$$\frac{\partial \tilde{u}}{\partial t} + \widetilde{R}(F(\tilde{u})) = \widetilde{R}(F(\tilde{u})) - \overline{R(F(u))} \qquad (3.2.1)$$

where $\tilde{u}$ represents exact solution for filtered velocity, $\widetilde{R}$ is a discrete representation of the spatial terms computed from $\tilde{u}$, $\overline{R(F(u))}$ denotes spatial terms (with a suitable closure for the unknown SGS stress) predicted by the residual NN. To increase the numerical stability, the authors reformulated the predicted term into data-informed (filtered) closure models of both the mixed and eddy viscosity type (the last one showed a better numerical stability).

**DNS / LES / knowledge / experiment to RANS**. RANS is one of the most popular approaches in CFD. It allows to consider moderate engineering scale applications. As a result, there are much more attempts to use high-fidelity data to correct errors in RANS simulations and, thereby, bridge the scaling gap caused by averaging of the NS equations.

Zhang & Duraisamy (2015) employed a multiscale Gaussian process (GP) to predict a multiplier $\alpha$ for the production term $\mathcal{P}$ in the equation for TKE $k$:

$$\frac{Dk}{Dt} = \alpha \cdot \mathcal{P} - \mathcal{D} + \mathcal{T} \qquad (3.2.2)$$

where $D/Dt$ denotes substantial derivative, $\mathcal{D}$ and $\mathcal{T}$ are destruction and transport terms, respectively. The multiplier $\alpha$ is introduced to reduce discrepancy between high-fidelity DNS data and results obtained with the $k-\omega$ model. Therefore, it serves as a model error correction term in the physics-based turbulence model. Similarly, Parish & Duraisamy (2016) used Bayesian field inversion to obtain corrective, spatially distributed term $\alpha(x,y)$ from DNS data to address model form errors in the $k-\omega$ model. Singh & Duraisamy (2016) inferred the posterior distribution for the multiplier $\alpha$ in the Spalart-Allmaras model:

$$\frac{D\tilde{\nu}_t}{Dt} = \alpha(x,y) \cdot \mathcal{P} - \mathcal{D} + \mathcal{T} \qquad (3.2.3)$$

where $\tilde{\nu}_t$ is surrogate viscosity. A key enabling idea of this approach is the transformation of the infinite-dimensional functional inversion procedure into a finite-dimensional problem in which the distribution of the unknown function is estimated at discrete mesh locations. The approach is demonstrated for a channel flow, shock-boundary layer interactions, and flows with curvature and separation. It is shown that even with limited data the accuracy of the inferred solution is improved in the entire computational domain. The drawback of using the spatially distributed corrector $\alpha(x,y)$ is that it can be obtained only for a specific geometry. Slightly different, paper (Duraisamy et al., 2015) suggested introducing a DD multiplier $\alpha$ for the destruction term:

$$\frac{D\tilde{\nu}_t}{Dt} = \mathcal{P} - \alpha \cdot \mathcal{D} + \mathcal{T} \tag{3.2.4}$$

or an additional term δ (a more generic approach) in the equations for eddy viscosity or TKE in the linear eddy viscosity models:

$$\frac{Dk}{Dt} = \mathcal{P} - \mathcal{D} + \mathcal{T} + \delta \tag{3.2.5}$$

The authors (Duraisamy et al., 2015) also demonstrated the performance of such error correction for the bypass transition problem (occurs when free-stream turbulence introduces instabilities in the evolving boundary layer to become turbulent):

$$\frac{D\gamma}{Dt} = \alpha + \frac{\partial}{\partial x_j}\left[(\sigma_l \nu + \sigma_\gamma \nu_t)\frac{\partial \gamma}{\partial x_j}\right] \tag{3.2.6}$$

where $\gamma$ is intermittency factor, $\sigma_l$ and $\sigma_\gamma$ are tunable parameters.

Wang et al. (2017) and Xiao et al. (2017) suggested a methodology aimed at improving RANS-modeled Reynolds stress in the $k - \varepsilon$ model by learning the functional form of the Reynolds stress discrepancy $\Delta R$ (their magnitudes and shape and orientation of the anisotropy) based on available DNS and LES data for flows in a square duct and for a periodic hills problem:

$$\frac{\partial U_i}{\partial t} + U_j \frac{\partial U}{\partial x_j} = -\frac{1}{\rho}\frac{\partial P}{\partial x_i} + \frac{\partial}{\partial x_j}\left[\nu\left(\frac{\partial U_i}{\partial x_j} + \frac{\partial U_j}{\partial x_i}\right) - (R_{ij} + \Delta R)\right] \tag{3.2.7}$$

To learn the discrepancy, a random forests (RF) model and a feedforward NN (FNN) were employed. The RF model provided physical insights regarding the relative importance of input features that contributed to the discrepancies. Selection of input features is an important issue for training of a ML algorithm. Irrelevant features can significantly deteriorate its performance, while absence of important features will undermine the training effectiveness. Since there is no *ad hoc* information on how to select these features, this problem should be analyzed thoroughly before using a DD approach. Similarly, Tan et al. (2020) explored an opportunity to directly correct errors in the turbulent viscosity $\Delta \nu_t$ for the Spalart-Allmaras model using a RF algorithm.

The subsequent work (Wang et al., 2016; 2018a) was aimed at the improvement of the predictive capability by expanding the input feature space by building an integrity basis of invariants, application for different turbulence models (*e.g.*, $k - \omega$ model) (Huang et al., 2017), as well as *a priori*[8] (Wang et al., 2019) and *a posteriori*[9] estimation of the framework's performance for different flow conditions.

Paper (Wu et al., 2017a) is focused on using of the distance metric in feature space between training and test flows to assess the data coverage condition. Specifically, the Mahalanobis distance and the kernel density estimation (KDE) technique are used to quantify the distance between flow datasets in feature space. The results show that the prediction error of the Reynolds stress anisotropy is positively correlated with Mahalanobis distance and KDE distance, demonstrating that both extrapolation metrics can be used to estimate the prediction confidence *a priori*. A quantitative comparison using correlation coefficients shows that the Mahalanobis distance is less accurate in estimating the prediction confidence than KDE distance. Wu et al. (2017b) used t-distributed stochastic neighbor embedding (t-SNE) methodology to reduce the dimensions of CFD datasets for improved visualization; such visualization technique enables easy similarity comparisons between datasets. Thus, one can see that data coverage assessment is also a crucial part of using ML for CFD. It is almost meaningless to use a ML model for prediction outside the training domain, but as discussed above, answer to the question whether training data sufficiently cover data for prediction is not trivial.

---

[8] In this work, *a priori* denotes high-dimensional visualization techniques, distance metrics, or other measures to assess the data coverage condition. Since ML algorithms work well in the interpolation regime, data coverage assessment provides merits to answer the question whether the test dataset is sufficiently covered by the training data.

[9] In this work, *a posteriori* denotes the assessment of a model after the training (by direct comparison of ML-predicted quantities with ground truth values and / or using the model during simulations). On the contrary, some authors by *a priori* denote direct comparison of ML-predicted quantities with ground truth values, while *a posteriori* is referred as usage of ML models during simulations.

Duraisamy et al. encouraged to use prior knowledge (*e.g.*, physical constraints for the Reynolds stress) and data for Bayesian inference to make parametric corrections to existing models and get direct information on the model inadequacy which is of value to the modeler. Ling et al. (2016a; 2016b) also pointed out importance of the physical constraints for the Reynolds stress (invariance properties and non-negativity of the eddy viscosity). The physical constraints are a good instrument to reduce the degree of freedom for the considered systems. Subsequently, in paper (Ling et al., 2017) a RF model (trained on LES data for flows in a duct and around a wall-mounted cube for the $k - \varepsilon$ model and tested on jet-in-crossflow configuration) was employed to predict barycentric coordinates. The coordinates are then used to modify the eigenvalues of the anisotropic part of the Reynolds stress to construct a more accurate Galilean invariant closure. The comparison is performed using *a posteriori* study, without online employment of the ML model for simulations.

In the subsequent work, Wu et al. (2019) demonstrated that RANS simulations with explicit DD Reynolds stress representation can be ill-conditioned. The authors suggested a metric based on local condition number function for *a priori* evaluation of the conditioning of the RANS equations. To overcome the issue of ill-conditioned RANS, systematic procedure to generate input features based on the integrity basis for mean flow tensors was developed (Wu et al., 2018). Inspired by a general form of eddy viscosity model (Pope, 1975), the anisotropic $a_{ij}$ and isotropic $\frac{2}{3}k\delta_{ij}$ parts of the Reynolds stress tensor were predicted separately by a RF model:

$$R_{ij} = a_{ij} - \frac{2}{3}k\delta_{ij} \tag{3.2.8}$$

Such a decomposition is instrumental in overcoming the ill-conditioning of the RANS equations. The RF model was trained on the DNS data for improvement of the $k - \varepsilon$ model for flows in a square duct and over periodic hills.

Xiao et al. (2016) introduced a DD, Bayesian approach for quantification and reduction of model form uncertainties in RANS simulations. Prior knowledge of physics (constraints on the Reynolds stress, such as physical realizability, spatial smoothness, and problem specific knowledge for a particular flow) is used for Bayesian inference-based calibration. Model form uncertainties in the RANS-predicted Reynolds stress tensor (TKE, anisotropy, and orientations in the $k - \varepsilon$ turbulence model) are obtained, which are then used for improvement of the simulation results for periodic hills and square duct flows. Wu et al. (2016) furtherly improved the framework making it applicable for flows without observed data by using Karhunen-Loeve expansion to build a surrogate for the Reynolds stress tensor. This allowed to develop a statistical model for the uncertainty distribution of the Reynolds stress discrepancy. The obtained distribution is then sampled to correct the RANS-modeled Reynolds stress for a flow to be predicted.

**RANS / experiment to CG / SYS TH**. SYS TH codes are designed to model the whole primary and / or secondary circuits of nuclear power plants. 1D lumped parameters SYS TH codes are the workhorse for SYS TH simulations, but they are gradually becoming legacy codes. Due to the increased computational power, 3D CG codes (including SYS TH codes with 3D capabilities) offer a promising substitute for them. It is natural to use 3D RANS simulations or experimental data to correct model / mesh induced errors in 3D CG codes.

Work by Bao et al. (2019) is directly focused on bridging the scaling gap between RANS and CG simulations. An FNN is trained to point-wise map physical quantities between the two "scales" and predict (i) model error (wall friction, turbulence, and convective heat transfer); (ii) mesh-induced error, which exists since the mesh size serves as a model parameter in closures for CG. The approach is called OMIS (optimal mesh / model information system) since it suggests optimal selection of computational mesh size (as a "computational" filter) and closures ("physical" filter). Additionally, the authors introduced 4 physics coverage conditions[10] and showed that the global extrapolation through local interpolation (GELI) (represents the situation when a global physical condition of a target case is an extrapolation of existing cases, but the local physics is similar) is an excellent area for ML to bridge the scaling gaps. The extrapolation of global physics indicates different global physical conditions such as a set of dimensionless parameters, or different initial and BCs, or different geometries, *etc.* The interpolation or similarity of local physics is dependent on the identification of physical features, data quality and quantity. The local similarity makes it feasible to derive great benefits from the existing data to estimate the target case. For example, local physics is similar for turbulent flows in round and square tubes; this similarity of local physics can be used to "transfer knowledge" from one case to the other. The performance of the developed approach is demonstrated for 2D mixed convection case study. In their subsequent work (Bao et al., 2020a), the authors extended the framework (working on input feature selection) to additionally account for scaling uncertainty due to the modified global conditions (geometry of the computational domain and BCs extrapolation). One of the difficulties of the approach is in the assessment of local physics similarity, which is methodologically similar to data coverage assessment procedure. In their work, Bao et al. used t-SNE method to reduce the dimensionality and visualize the data.

Torres-Rua et al. (2012) successfully used data from an irrigation conveyance canal to train an FNN and support vector machine (SVM) models for prediction of errors in hydraulic simulation model (can be viewed as an analog of SYS TH codes in the nuclear industry). Zhao et al. (2020) also trained an FNN and RF models on experimental data to predict errors in a physics-based model for critical heat flux prediction. They found out that such hybrid framework outperforms a standalone DD closure model.

*3.2.2. Computational cost reduction approaches*

As it is shown above, DD error correction is successfully being used as the high-to-low methodology. At the same time, there are examples when it is employed to reduce the computational cost of high-fidelity tools. Below some of these approaches are discussed.

**Grid coarsening**. Hanna et al. (2020) suggested a framework for CG DNS. FNN and RF models were separately trained to predict the discretization errors based on the fine mesh data for the flow in 3D lid-driven cavity. The proposed approach is capable to correct the CG results for new cases (with different Reynolds numbers, or different grid spacings, or different aspect ratios of the cavity). Similarly, Bao et al. (2020b; 2021) suggested to use an FNN to map high-fidelity two-phase RANS simulations with CG RANS simulations and thereby predict the discretization error. The workflow is demonstrated for a bubbly flow case study: unphysical patterns (peaks in the velocity and void fraction profiles near the wall) in the CG simulation were successfully captured and corrected. Zhu et al. (2020) developed a framework to map low-fidelity velocity with its high-fidelity value thereby correcting errors caused by insufficiently small mesh size. The authors mention that direct prediction of errors can cause numerical instability since the magnitude of changes of errors is relatively high. The approach showed the potential to save the computational time of RANS simulations by making the mesh coarser.

---

[10] Physics coverage condition has a slightly broader meaning than the data coverage condition.

Another interesting approach to use CG CFD is based on the superresolution technique (Brunton et al., 2020), which takes its roots from the computer vision area. Such algorithms are trained to predict high-fidelity fields from low fidelity "images" thereby showing potential in reconstructing sparse data. Fakami et al. (2019) showed the ability of CNNs to reconstruct turbulent flows from extremely coarse fields. Kim et al. (2020) presented an unsupervised ML model (which not requires labeled high-fidelity target data) that adopts a cycle-consistent generative adversarial network. The performance of the model is demonstrated for filtered DNS data reconstruction, and for LES data reconstruction to DNS quality. Paper (Xie et al., 2018) investigates the performance of a temporally coherent generative model addressing the superresolution problem for 3D fluid flows. Similar approaches are also investigated in (King et al., 2018).

**Grid optimization**. To predict discretization error in simulations and enable online mesh adaptation, Chen & Fidkowski (2020) suggested a DD approach (based on a CNN) for adjoint-based error estimation (no high-fidelity data is needed) for steady state Burger's equation solver. Adjoint-based methods provide an approach to quantify the output error and to guide the mesh adaptation, but they require an adjoint solution, which imposes both implementation and cost challenges in practice. The authors demonstrated that ML methods can leverage this process.

**Initial approximation and co-simulation**. This methodology is classified as "correcting errors in physics-based models" since technically it is aimed at correcting "wrong" initialization to get a faster convergence to steady state. To reduce the computational cost of simulations, Guo et al. (2018) used CNNs for initial approximation of steady state CFD simulations. Hodges et al. (2019) also used CNN (transposed) to predict temperature distribution during fires in simple geometries. They showed that using BCs as inputs to the CNN, 2D temperature distributions can be predicted with satisfactory errors. To extrapolate the predictions for other geometries, the authors tried to identify input features which can provide the ML algorithm with additional information about the global physical condition. Similarly, Sekar et al. (2019) used a CNN to predict flow around an airfoil; in addition to the geometry of airfoils, the authors used Re number and attack angle to improve prediction accuracy. Paper (Ye et al., 2020) adopts a similar strategy to predict pressure distribution around a cylinder.

Inspired by the traditional co-simulation methodology (*e.g.*, see (Barbosa & Mendes, 2008)), Mazuroski et al. (2018) used ML to enable RANS simulations for flow around buildings. The authors used recurrent NN ("prediction" model) to approximate the current state in time and then use this information to obtain a more accurate solution using a CFD-tool ("complex" model).

## 3.3. Turbulence Models

Classification of DD turbulence models is shown in Fig. 3.4. Among high-to-low approaches following are distinguished: (1) informing LES by DNS and lattice Boltzmann methods (LBM); (2) informing RANS simulations by DNS / LES / knowledgebase, and/or experimental data; (3) informing CG or SYS TH simulations by DNS or RANS simulations. Computational cost reduction approaches can be subdivided into (1) flow reconstruction for regions with high mesh resolution requirements; and (2) surrogate modeling, when a computationally efficient DD model is developed to mimic a physics-based model.

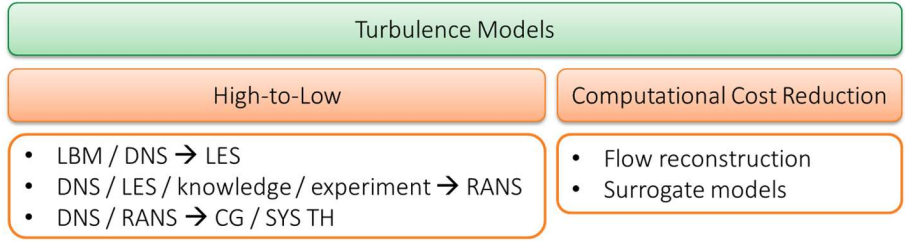

Fig. 3.4. Classification of DD approaches for turbulence modeling.

### 3.3.1. High-to-low approaches

**LBM / DNS to LES**. ML methods can benefit from abundance of high-fidelity data by building turbulence models for CFD including LES. Wang et al. (2018b) trained an FNN and RF models to build a SGS stress model for LES using DNS data. Symmetric SGS stress tensor was predicted based on a set of input features for which a careful relative importance analysis was performed. It is shown that the gradient of filtered velocity and the second derivatives of filtered velocity account for a vast majority of the importance. More importantly, they claim that different SGS stress components have different dependences on the input features and there exists a specific pattern. Performance of the model is analyzed for isotropic turbulence. In a similar way, Zhou et al. (2019) used an FNN to predict SGS stress for LES. The Gaussian filtered velocity gradients tensor obtained in DNS were used as input features. *A posteriori* test (correlation coefficients between DNS and FNN predictions and employment of the model for LES for isotropic turbulence) has shown that it outperforms the conventional models (the gradient model, the Smagorinsky model, and its dynamic version). To compare the results the authors used the Lagrangian statistics of fluid particle pairs (mean and variance of the separation distance, relative dispersion, and Lagrangian velocity correlations).

Maulik et al. (2018) used an FNN for convolution and deconvolution (filters) of fields to account for SGS turbulence effects in LES (predict effective viscosity from DNS data). The proposed framework is also combined with a statistical truncation mechanism for ensuring numerical realizability of an explicit formulation. Performance of the framework is demonstrated for a 2D decaying turbulence. It is shown that a robust and stable SGS closure can be generated, which compares favorably to the Smagorinsky and Leith hypotheses for modeling of the Kraichnan turbulence. In (Chen, et al., 2019) data from DNS of the atmospheric boundary layer are employed to build a deep learning model for LES SGS stress. The model showed higher accuracy comparing to the Smagorinsky and the Smagorinsky-Bardina mixed models. The SGS stress at each spatial point is predicted using resolved velocity in the neighboring box in a similar way to convolutional operation using an FNN. The authors tested different sizes of the convolution kernel ($7 \times 7 \times 7$, 5×5×5, and 3×3×3) and found out that there is not significant difference between them, which allows to say that stress essentially depends on the local neighboring velocity field. Among useful input features, pressure and temperature are employed to increase the accuracy of the model.

Xie et al. (2020) reconstructed SGS stress and SGS heat flux using an FNN for compressible turbulence. The input features for the NN are the first- and second-order derivatives of filtered velocity and temperature from DNS at different spatial locations. In addition, a thirteen-point stencil is designed to model the effect of spatial structures of flow field at the scales near the filter width. The stencil is determined by the grid length in each dimension and due to that the authors called the NN a "spatial NN".

Wang et al. (2020) introduced trainable spectral filters in a coupled model of RANS and LES. The hybrid RANS-LES approach uses three-level decomposition of the velocity field:

$$u = U + \tilde{u} + u' \qquad (3.3.1)$$

where $u$ is instantaneous velocity, $U$ is Reynolds-averaged component, $\tilde{u}$ is filtered component, $u'$ is fluctuating component. The velocity field is decomposed into three components of different scales using two scale separation operators (spatial and temporal filters). In traditional CFD, these filters are usually pre-defined, such as the Gaussian spatial filter. In the introduced TF-Net (turbulent flow network), both filters are trainable NNs. The spatial filtering process is instantiated as a one-layer CNN with a single 5×5 filter to each input image. The temporal filter is also implemented as a convolutional layer with a single 1×1 filter applied to several successive images. The TF-Net is trained on 2D LBM data. The authors provided exhaustive comparisons of TF-Net with other types of NNs and observed significant improvements for prediction error and desired physical quantifies, including divergence, TKE, and energy spectrum. Lav et al. (2019) also employed three-level decomposition Eq. (3.3.1). DNS data were used to predict anisotropy tensor for RANS equations, unsteady RANS (URANS), and partially averaged NS (PANS) equations. This new method achieves improvement by resolving the length scales that are responsible for the organized unsteadiness for RANS and modeling the remaining length scales. A case study, of a normal flat plate was considered to demonstrate the applicability of the methodology. A non-linear turbulence closure is then developed based on the "unsteady" part of the DNS anisotropy. Using this closure for URANS simulations of the same case study a significant improvement in the mean flow statistics compared with the baseline URANS is shown. Further improvements were obtained when the turbulence equations were converted from URANS to PANS.

Yang et al. (2019) argued that SGS modeling in LES usually needs flow information only at large scales, in contrast with wall modeling, which needs to account for both near-wall small scales and large scales above the wall. They addressed this challenge by using spatially filtered DNS data and an FNN to predict wall shear stress in LES. To improve the extrapolation capabilities of the FNN, the authors considered using knowledge-based feature normalization.

**DNS / LES / knowledge / experiment to RANS**. High-fidelity data and knowledgebase are successfully being used for building turbulence closures for RANS. Thus, Zhang & Duraisamy (2015) used experimental wall shear stress data to build a multiscale GP and FNN-based models for RANS simulations to substitute production and destruction terms in the Bypass transition model Eq. (3.2.6). The aim of the multiscale GP (Zhang et al., 2016) is to ensure accuracy in situations where the parameter space contains regions that are sparsely and densely populated. The multiscale GP features altered radial basis functions ϕ at each spatial location (*i.e.*, $h_n$ depends on a spatial location):

$$\phi(q_m, q_n) = e^{-\frac{\|q_m - q_n\|^z}{h_n^2}} \tag{3.3.2}$$

Comparing to the conventional GP and FNN, such novel approach showed a better performance by employing DNS data for modeling 1D channel flow and employing experimental data for the bypass transition phenomenon.

Ling et al. (2016a) is proposed to use a more general formulation of eddy viscosity hypothesis (Pope, 1975). By application of the Cayley-Hamilton theorem, it can be shown that every second-order anisotropic tensor can be expressed in the following form:

$$a_{ij} = \sum_{\lambda=1}^{10} G^\lambda(I_{ij}^{1:5}) \cdot T_{ij}^\lambda \tag{3.3.3}$$

where $T_{ij}^{1:10}$ are tensor basis functions, $G^{1:10}$ are tensor basis coefficients which depend on the invariants $I_{ij}^{1:5}$:

$$I_{ij}^{1:5} = [\{s_{ij}^2\};\ \{\omega_{ij}^2\};\ \{s_{ij}^3\};\ \{\omega_{ij}^2 s_{ij}\};\ \{\omega_{ij}^2 s_{ij}^2\}] \tag{3.3.4}$$

$$\begin{aligned}
T_{ij}^{1:10} = \Big\{ & s_{ij};\quad s_{ij}\omega_{ij} - \omega_{ij}s_{ij};\quad s_{ij}^2 - \frac{1}{3}\delta_{ij}\{s_{ij}^2\}; \\
& \omega_{ij}^2 - \frac{1}{3}\delta_{ij}\{\omega_{ij}^2\};\quad \omega_{ij}s_{ij}^2 - s_{ij}^2\omega_{ij};\quad \omega_{ij}^2 s_{ij} + s_{ij}\omega_{ij}^2 - \frac{2}{3}\delta_{ij}\{s_{ij}\omega_{ij}^2\}; \\
& \omega_{ij}s_{ij}\omega_{ij}^2 - \omega_{ij}^2 s_{ij}\omega_{ij};\quad s_{ij}\omega_{ij}s_{ij}^2 - s_{ij}^2\omega_{ij}s_{ij};\quad \omega_{ij}^2 s_{ij}^2 + s_{ij}^2\omega_{ij}^2 - \frac{2}{3}\delta_{ij}\{s_{ij}^2\omega_{ij}^2\}; \\
& \omega_{ij}s_{ij}^2\omega_{ij}^2 - \omega_{ij}^2 s_{ij}^2\omega_{ij} \Big\}
\end{aligned} \tag{3.3.5}$$

where the mean rate-of-strain tensor (symmetric tensor)

$$s_{ij} = \frac{1}{2}\frac{k}{\varepsilon}\left(\frac{\partial U_i}{\partial x_j} + \frac{\partial U_j}{\partial x_i}\right) \tag{3.3.6}$$

and the rotation tensor (antisymmetric tensor)

$$\omega_{ij} = \frac{1}{2}\frac{k}{\varepsilon}\left(\frac{\partial U_i}{\partial x_j} - \frac{\partial U_j}{\partial x_i}\right) \tag{3.3.7}$$

are both normalized by TKE $k$ and turbulent dissipation rate $\varepsilon$. Therefore, the invariant NN is designed to learn tensor basis coefficients $G^{1:10}$ to produce the anisotropic component $a_{ij}$ from invariants $I_{ij}^{1:5}$ and tensor basis functions $T_{ij}^{1:10}$. To test the invariant NN, DNS and well-resolved LES data are used to predict the anisotropic part in RANS solver with the $k - \varepsilon$ model. The invariant NN showed ability to be used for extrapolation cases, a better performance than the standalone $k - \varepsilon$ model, and a NN without invariant properties. Paper (Geneva & Zabaras, 2019) uses a Bayesian formulation of the invariant NN. Instead of predicting a single value for the anisotropic term, the authors predict a sample of values. Based on that information, subsequent uncertainty propagation to quantities of interest (velocities and pressure) is performed. Such approach not only allowed to improve the prediction of the $k - \varepsilon$ model, but also quantify model form uncertainties. It is worth to note that the authors used completely different sets of training and testing flows. Even though they observed a noticeable improvement of the RANS simulations comparing to LES with the Smagorinsky model, further improvements are required to deal with complex phenomena in flows such as vortices behind objects and flow separation.

Paper (Ling et al., 2016b) serves as a further exploration of embedding invariant properties in DD turbulence models. The authors used DNS data to improve RANS simulations with the $k-\varepsilon$ model for a wall-mounted cube problem. RF and FNN were trained by different methods to embed the invariance. In the first method, a basis of invariant inputs is constructed (preprocessed data), and the ML model is trained upon this basis, thereby embedding the invariance into the model (similarly to (Ling et al., 2016a)). In the second method, the algorithm is trained on multiple transformations of the raw input data until the model learns invariance to that transformation. It is shown that embedding the invariance property into the input features yields better performance at significantly reduced computational training costs.

Taghizadeh et al. (2020) furtherly proposed and investigated three elements to ensure the physical consistency of ML turbulence closures, namely: (i) characteristic physical features and constraints that all (physics-based and ML) closure models must strive to satisfy; (ii) ML training scheme that infuses and preserves selected physical constraints; and (iii) physics-guided cost function to optimize model's predictions. Therefore, additionally to the invariant NN (Ling et al., 2016a), the authors introduced a physics-informed cost function:

$$\mathcal{L} = \frac{1}{4N}\sum_{m=1}^{N}\left[\sum_{k=1}^{3}(a_{kk}-\hat{a}_{kk})^2 + \frac{1}{u_\tau^4}\left(\langle u'_1 u'_2\rangle - \widehat{\langle u'_1 u'_2\rangle}\right)^2\right] \qquad (3.3.8)$$

where predictions by the NN are denoted as $\hat{a}_{kk}$ (normalized anisotropic component) and $\widehat{\langle u'_1 u'_2\rangle}$, while the true values from DNS are $a_{kk}$ and $\langle u'_1 u'_2\rangle$; $N$ is the number of data points, $u_\tau$ is friction velocity from DNS (for normalization). Additionally, the authors introduced an iterative framework to embed physics-based constraint to the model coefficients (calibration on the canonical cases: equilibrium behavior of homogeneous turbulence and equilibrium behavior of log-layer). The proposed approach is investigated in a simple channel flow. In the evaluation process, the standard $k-\omega$ model is intentionally altered from its original form and the ability of open loop and closed loop frameworks to embed the physical constraints is examined.

Another approach to implement physical constraints in a DD turbulence model is explored in (Fang et al., 2019). The authors used an FNN with special architecture to predict anisotropic term with enforced BCs, Re number, and spatial non-locality effects. DNS data are used to build a model for RANS with linear viscosity model for fully turbulent flow in a channel. As a first step, a generic FNN was used as the base model, and then its capabilities were increased by using non-local features, directly incorporating Re number (information on the bulk flow) and enforcing the BCs at the channel wall by predicting velocity gradient near the walls. The models are trained and tested on different combinations of Re numbers in order to investigate the generalizability. The models are compared with the invariant NN introduced in (Ling et al., 2016a). The comparison showed that the introduced models outperformed the invariant NN for the considered scenario.

Additionally to the physics-informed cost function, Raissi et al. (2019) used deep hidden physics models to predict unclosed terms for turbulent scalar mixing. Automatic differentiation is used to compute the derivatives and build the residual (physics-informed) NN. Sparse high-fidelity experimental measurements are used to "implicitly" improve diffusion and dissipation for turbulent scalar mixing. Similar PDE-informed NN is designed to extract eddy viscosity and TKE, derivatives of the Reynolds stress from DNS data in (Iskhakov & Dinh, 2020a; 2020b). The DD turbulence models were learnt directly from velocity and pressure fields. This is enabled by informing the NN by numerical solution of the NS equations.

Weatheritt & Sandberg (2017) symbolically regressed algebraic forms of the Reynolds stress anisotropy tensor Eq. (3.3.3) for the SST $k-\omega$ turbulence model from hybrid RANS-LES data. Input and target anisotropy tensors are calculated from the "frozen" simulation data (the equations are solved for fixed velocity and Reynolds stress for "correct" value of ω). ML models that follow equation Eq. (3.3.3) can be explicitly provided through symbolic regression by (evolutionary) gene-expression programming (GEP), which is used to determine the tensor basis coefficients $G^{1:10}$ based on training data. The framework is computationally inexpensive and produce accurate and robust models *with available mathematical expressions*. This interpretability of the results allows to diagnose issues with the regressed expressions and correct them if necessary. RANS solver with GEP-trained models shows improved predictive accuracy for rectangular ducts (Weatheritt & Sandberg, 2017) and turbomachinery flows (Akolekar, et al., 2019). Later Zhao et al. (2020) extended this framework by integrating the GEP with RANS solvers, so the cost function did not connect anisotropy tensors, but any flow feature from simulations. This approach is ideologically similar to (Iskhakov & Dinh, 2020a; 2020b), but more practical (does not involve development of a new solver) and, in some cases, can be more computationally expensive due to its iterative nature.

Zhu & Dinh (2020) extracted Reynolds stress from DNS data (fully developed incompressible pressure-driven turbulent flow between two parallel planes), which are considered as physically correct data (targets). Flow features are extracted from RANS results with the $k-\varepsilon$ model (inputs). An FNN is trained to map the inputs and targets. In the subsequent work (Zhu et al., 2020), the authors employed an ensemble (multi-model) learning technique for a complex flow configuration (3D sub-channel of a pressurized water reactor) to map DNS data with RANS $k-\varepsilon$ model / RANS $k-\omega$ model / laminar flow data to predict samples of the Reynolds stress for RANS solver. Such multi-model DD turbulence modeling shows higher accuracy than both traditional turbulence models and single-model ML frameworks.

**DNS / RANS to CG / SYS TH**. To bridge the scaling gap between RANS and CG TH simulations (utilize fine mesh RANS data to develop a CG DD turbulence model), Liu et al. (2020) used a 3D densely-connected CNN to predict turbulent viscosity field from velocity, pressure, and temperature fields from CG data. The transient used as a case study involves thermal mixing and stratification (loss-of-flow accident in a metal-cooled reactor core), a challenging problem for SYS TH simulations with complex TH behavior. The framework showed a good potential; however, there were some difficulties caused by transient nature of the flow.

Gairola et al (2019) used statistical mechanics principles (Langevin equation-based statistical surrogates (Pope, 2011; Risken, 1996; Friedrich & Peinke, 1997)) to predict velocity and passively advected scalar quantity (temperature in liquid metal) by relying on the Lagrangian description of the flow field. The drift and diffusion coefficients of the Langevin equation were obtained with help of Kramers-Moyal expansion, capturing the Markovian processes from DNS data. The resulting emulator can be coupled with a SYS TH code for tracking the position of a particle which can subsequently be used to estimate the scalar field fluctuations or mixing extent under turbulent flow conditions. In other words, the developed statistical surrogate emulated turbulent velocity signal with high degree of accuracy.

Mak et al. (2018) introduced common proper orthogonal decomposition (CPOD) for a swirl injector performance prediction. Such CPOD can be viewed as a physics-guided partition of the computational domain to allow POD (which is only suitable for extracting instability structures at a single geometry) into temporal and spatial components:

$$Y(x,t) = \sum_{k=1}^{\infty} \beta_k(t)\phi_k(x) \qquad (3.3.9)$$

where $Y$ is a flow variable, $x$ is spatial coordinate. CPOD extracts common instabilities over the design space in time. At each simulation time-step a sparse covariance matrix is employed to account for the few significant couplings among flow variables. The emulator provides accurate flow predictions and captures several key metrics for injector performance from high-fidelity simulations. The proposed model offers two appealing features: (a) it provides a physically meaningful quantification of spatial and temporal uncertainty, and (b) it extracts significant couplings between flow instabilities.

### *3.3.2. Computational cost reduction approaches*

The aim of such approaches is to reduce the computational cost of the existing physics-based models. The aim of improving the predictive capability is not considered.

**Flow reconstruction**. Milano & Koumoutsakos (2002) were among the first researchers who used ML (linear POD and an FNN (non-linear POD)) for reconstruction of the channel flow quantities (wall-normal velocity, streamwise, and spanwise velocities) near the wall. They used DNS data to teach a ML algorithm to reconstruct the flow near the walls where the requirements for grid resolution are especially high. The ML model requires the entire field as input, producing its compressed version. In a similar manner, paper (Bourguignon et al., 2014) describes an approach to use velocity fields from DNS to reconstruct streamwise velocity near the walls by using successive Fourier-transform and compressive sampling. This allowed to significantly reduce computational cost of the subsequent DNS for canonical flows in channels.

**Surrogate models**. Paper (Sarghini et al., 2003) investigates opportunity to reduce computational time by building a surrogate SGS stress model for LES. An FNN is trained on LES Bardina's scale similar SGS model. Among the input features, the authors chose 9 velocity components, and 6 components of the stress tensor. The model successfully worked only in a limited range of Re numbers (in interpolation mode).

Tracey et al. (2015) investigated the feasibility of building an FNN-based turbulence model (eddy viscosity) by attempting to reproduce the Spalart-Allmaras model. The trained NN successfully reproduced it in a wide variety of flow conditions from 2D flat plate boundary layers to 3D transonic wings, including flow conditions previously unseen by the model. Similarly, in (Sun et al., 2019) and (Zhu et al., 2019a) the authors trained FNNs as surrogate substitutes to the Spalart-Allmaras model. They used optimal brain surgeon (Augasta & Kathirvalavakumar, 2013) to determine the relevancy of input features, which boils down to the sensitivity analysis of weights of input neurons.

"An empirical approach" was developed in (Chang et al., 2020). The authors used feature coverage mapping based on the t-SNE to quantify the data coverage condition of ML-based closures. A backward-facing step flow is used to demonstrate that not only can NNs discover underlying correlation behind fluid data but also, they can be implemented in RANS to predict flow characteristics without numerical stability issues. Training data are generated from RANS simulations with the $k-\varepsilon$ model. Quantity to be predicted by the ML model is Reynolds stress from input spatial derivatives of velocity fields. Closure relations are iteratively queried from the DD model when solving conservation equations.

Zhu et al. (2019b) successfully built an FNN-based eddy viscosity model for 3D thermal stratification problem, which was modeled by a CG code. The authors also discuss the problem of imbalanced training datasets and note that it can be tackled by over-sampling and under-sampling techniques. However, to simplify the workflow, the authors split all data into several datasets and monitored the performance of the ML model by training on the reduced sets.

### 3.4. Two-phase Flow and Thermal Fluid Closures

Two-phase flow and thermal fluid modeling with RANS and SYS TH require a bunch of closures (Kulacki, 2018; Liu et al., 2017) including, but not limited to: convective, boiling, condensation heat transfer, interfacial drag, lift, and other forces, *etc*. The aim of this subsection is not to give a detailed review (which can be found in (Cong et al., 2013), (Chang & Dinh, 2019), and (Gomes-Fernandez et al., 2020)), but rather to show that there are many attempts in the nuclear community to use ML to tackle the closure problem. Historically, physics-based closures are being developed in an attempt to fit the understanding of physics and experimental data. ML promise a higher potential for assimilation of data since it is not limited by a particular functional form, therefore, it has a higher potential to reduce model form uncertainties in simulations.

One of the closures needed for SYS TH simulations are two-phase flow regime maps. Papers (Yang et al., 2017) and (Du et al., 2019) consider an opportunity for building the maps using CNNs (classification problem). They use experimental data (photographs) of different flow regimes in round plexiglas tubes, though it is not clear how this approach can be applied in practice. CNNs automatically extract necessary features from pictures to differentiate bubbly, slug, and other flow regimes. SYS TH analysis requires classification based on other features, such as pressure, void fraction, and / or mass flow rates. From this point of view, paper (Hernandez et al., 2019) is more practical: the authors proposed a DD methodology for selecting closure relationships based on a decision tree model. As inputs dimensionless numbers were used. The closure laws selection model achieved high accuracy in classifying flow regimes for a wide range of two-phase flow conditions. Similar critique is applicable to Sierra et al. (2020) work, where the authors used experimental data (images captured from an experimental two-phase flow circuit) to train an FNN with randomized Hough transform to predict void fraction in two-phase flow.

Jung et al. (2020) used experimental data to build an FNN-based model which predict bubble size for two-fluid model. The model is useful for two-phase RANS simulations since the bubble size is a crucial parameter for interfacial closure laws. The authors also examined importance of the input features using sensitivity analysis, principal component analysis (PCA), and RF. Similarly, Ma et al. (2015) used DNS of bubbly flows to train an FNN for gas-phase flow rate and the streaming stresses prediction. The model serves as a closure for the average two-fluid equations.

Another important classification problem in SYS TH is boiling regime detection. Paper (Hobold & da Silva, 2019) introduces Bayesian statistics with CNNs for prediction of the transition from nucleate to film boiling regime (continuous vapor film formation). The CNN is trained on the experimental data and can predict the boiling crisis faster than temperature-based prediction systems with more than 99% accuracy. Again, we note that more practical approach would require manual feature extraction rather than usage of CNNs. Kim et al. (2015) used a cascade fuzzy NN to predict departure from nucleate boiling ratio. The data were generated using a coupled fuel performance and subchannel codes. Similar models may be useful for SYS TH simulations or reactor monitoring systems. Another important parameter during boiling crisis is wall temperature at critical heat flux. Park et al. (2020) considered an opportunity to build a surrogate FNN-based model for a SYS TH code to predict it. They showed that the ML model significantly reduced the computational time and had low discrepancy with the original physics-based model.

Heat transfer coefficient closures for different conditions are crucial for SYS TH simulations. Based on the experimental data Liang et al. (2020) performed *a priori* thermal performance evaluation of heat pipes by predicting Kutateladze number using NNs and genetic algorithm. Longo et al. (2020) developed an FNN-based closure to predict refrigerant condensation heat transfer coefficients in heat exchangers using experimental data. Seo et al. (2020) used DNS data for prediction of natural convection heat transfer (Nu number) from a long rectangular enclosure containing a sinusoidal cylinder. Qiu et al. (2020) trained an FNN on experimental data to predict flow boiling heat transfer coefficient in mini / micro-channels. The model showed a higher accuracy than traditional semi-empirical correlations. DD modeling for boiling heat transfer is also considered in (Liu et al., 2018). The authors thoroughly extracted features from high-fidelity pool boiling DNS to train an FNN for heat flux, void fraction, and wall superheat prediction. The results demonstrated that ML can be a promising tool to help improve the predictive capability of two-phase RANS solvers.

Accurate modeling of critical flows, which occur during initial stages of LOCA when a liquid under high pressure outflows into a low-pressure volume, is also a big challenge. Historically, several semi-empirical models were developed and used for such purposes (Elias & Lellouche, 1994). ML algorithms have a potential to outperform physics-based models for this problem due to their ability to map highly non-linearly depending parameters (pressure, subcooling, void fraction, type of a rupture, *etc.*). For such purposes, An et al. (2020) developed a framework based on a cascade fuzzy NN to substitute the well-known Henry-Fauske model and predict critical flowrate and pressure. It is reported that the model is highly accurate and much faster than the original physics-based one (it is not required to solve non-linear equations and look up the steam tables).

For simulation of beyond design-basis accidents with core meltdown it is important to predict the minimum film boiling temperature. Bahman & Ebrahim (2020) contributed by performing a review and collection of experimental data on the minimum film boiling temperature for various substrate rods quenched in high- and low-pressure distilled water pools. They used an FNN to synthesize the data by building a surrogate model, which can be potentially used in SYS TH simulations to replace conventional semi-empirical models.

Based on RANS simulations with the employed $k - \varepsilon$ model, Grbcic et al (2020) designed FNN and SVM-based models to predict parameters of a fluid after mixing in junctions. In a similar way, phase separation in junctions (Yang, et al., 2016) can be predicted by ML algorithms.

### 3.5. Facilitation of the Concurrent Modeling Approaches

The HMM is based on the compression and reconstruction operators Eq. (2.3.3), which allow connecting the micro- and macro-scales during simulations. As it is noted, design of such operations is an ill-posed and very difficult problem. Even though there are not many papers published which try to do so for the HMM, the usage of ML algorithms such as encoders, decoders, unsupervised algorithms, *etc.* is essential for such operations. For example, (Wang et al., 2020) used a NN for upscaling procedure. Thereby the authors obtained better approximations of the micro-model, which takes into account the observed data. The NN provided a nonlinear mapping between the time steps. The multiscale concepts, used in the NN, provide appropriate CG variables, their connectivity information, and some information about the mapping. Stephenson et al. (2018) used GP regression to predict micro-scale information based purely on macro-information, thereby avoiding costly repeated simulations of similar atomistic configurations. The EFM also heavily relies on the mappings between the micro- and macro-models, see Eqs. (2.3.4) – (2.3.7). The method also requires data compression and reconstruction (restriction and lifting operators). Potentially, DD-based operators can also ease the process for developing of such frameworks (E et al., 2020). One of the issues in the HMM and EFM that can be tackled by ML is the choice of time steps (Kuhl, 2019; Han et al., 2018) for temporal coupling.

ML can also assist in learning non-linear and / or stochastic PDEs (*e.g.*, (Raissi & Karniadakis, 2018)). Such computationally efficient models can represent missing micro- or macro-models (Yang et al., 2020; Xiao, et al., 2020; Chan & Elsheikh, 2018).

There is a high potential for using ML for the multigrid methods. For example, He & Xu (2019) demonstrated that there is a close connection between CNNs and multigrid methods. They showed that pooling and feature extraction in CNN correspond directly to restriction operation and iterative smoothers in the multigrid methods, respectively.

In adaptive mesh and algorithm refinement approaches, ML can be used either for better PDEs discretization (Bar-Sinai et al., 2019), classification purposes (*e.g.*, prediction of regions with high uncertainties (Ling & Templeton, 2015), and mesh refinement or switching to a more appropriate closure (Zhu et al., 2019a).

## 4. CONCLUSIONS

A literature review of physics-based and DD multiscale modeling methods is performed. Emphasis is placed on engineering scale applications, including (but not limited to) CFD and nuclear TH. A comprehensive introduction to the physics-based multiscale modeling methods is provided with highlights of notable successes, and discussion of assumptions and limitations of each of the approaches. Following the existing classifications, serial and concurrent modeling approaches are distinguished.

The serial modeling is subdivided into the approach based on pre-computed values and closure modeling. Large complexity of CFD and TH problems does not allow to use the pre-computation to take into account the multiscale nature of the fluid flows (though there are exceptions that prove the rule). Furthermore, it is infeasible to develop equation-based generic closures that will be applicable for a wide range of problems. This issue is discussed for closure problems in CFD and nuclear TH: turbulence models and two-phase and thermal fluid closures.

The concurrent modeling is subdivided into the multidomain approaches, HMM, and EFM with discussion of possible time coupling approaches. Even though the coupling of codes and models is an algorithmically (not only) challenging task, such methodologies are more promising for multiscale modeling. The multidomain methodology is one of the most popular approaches in the nuclear community (since historically many efforts are put to develop and validate codes; it is a natural step to couple them). Perhaps the major drawback of the multidomain methods is that both micro- and macro-models are considered to be complete. Therefore, the method heavily relies on the *ad-hoc* choice of closure models. As a result, it somewhat inherits drawbacks of the closure modeling approach. The HMM and EFM are free from this drawback, but additional level of complexity exists due to the necessity to develop compression and reconstruction (and other) operators and time coupling schemes to maintain correct information flow between the considered scales. Unfortunately, there are limited number of papers that apply HMM and EFM for engineering scale CFD and TH problems.

Additionally, the adaptive mesh & algorithm (model) refinement and the multigrid methods are briefly reviewed. Despite well-developed mathematical apparatus and ability to substantially reduce computational cost, the methods lack scalability for engineering and system scale problems (modeling of long transients with high mesh resolution may be infeasible).

A literature review of DD approaches to enable and / or improve multiscale CFD and TH simulations allowed to introduce a classification based on the goals for ML algorithms. It is shown that they can serve as (i) error correctors for physics-based models, (ii) standalone turbulent models, (iii) two-phase and thermal fluid closures, and (iv) part of the existing concurrent modeling approaches to facilitate them. The first three approaches can be furtherly subdivided into (a) high-to-low and (b) computational cost reduction methodologies.

While there are several opportunities for ML algorithms to enable / improve the multiscale modeling, there are many challenges associated with their application. One of the major issues is that (similarly to physics-based models) ML models are mostly applicable only within the training domain (interpolation). Thus, one has to carefully examine the data coverage condition – whether the training data sufficiently cover a target application (data quantity issue). Additionally, the training dataset can have a poor quality (*e.g.*, imbalanced, noisy). Working with experimental data, one will face the issue of data sparsity and incompleteness. All this requires careful activities related to data assessment and preprocessing.

Closely related to the data preprocessing is input features selection procedure. As demonstrated, there is no agreement on how to appropriately choose them (*e.g.*, some papers claim that second derivatives of velocity are not important for turbulent closures, while other papers claim otherwise). There are several attempts to deal with that issue: input features can be selected using iterative frameworks; some papers employ special techniques and algorithms to assess the relative input feature importance.

Furthermore, most of the ML models are black box ones and, therefore, lack of explainability. Even though, it is a rather philosophical question whether physics-based models are explainable, there are some efforts to move to more explainable ML models.

Numerical stability and convergence are big issues for employment of ML for fluid flows modeling (especially for high Re numbers). Physics-based models provide "smooth" solutions, while DD approaches may give an oscillating nature of the convergence. From that point of view, it is important to make ML closures physically consistent. This will not only allow to improve their predictive capability and decrease the degree of freedom, but also improve the generalization capabilities. Thus, incorporation of physical constraints (invariance, non-negativity of eddy viscosity, *etc.*) is a popular area for investigation.

Last, but not least, DD verification, validation, and uncertainty quantification (VV&UQ) activities for physics-based models is an instrument to improve the multiscale modeling (not considered in this paper). ML and statistical methods are powerful tools for calibration and VV&UQ analysis. Not to mention, UQ of ML models themselves is an active area of investigation nowadays.